\documentclass[a4paper,aps,prd,superscriptaddress,nofootinbib,notitlepage,twocolumn,floatfix]{revtex4-1}

\usepackage{graphicx}
\usepackage{amssymb,amsmath,amsfonts}

\usepackage{dcolumn}
\usepackage{bm}
\usepackage{color}
\usepackage{dsfont}
\usepackage{mathrsfs}
\usepackage{mathtools}
\usepackage{multirow}
\usepackage{slashed}
\usepackage{bbold}
\usepackage{physics}
\usepackage{cancel}

\newcommand \be{\begin{equation}}
\newcommand \ee{\end{equation}}
\newcommand \bea{\begin{eqnarray}}
\newcommand \eea{\end{eqnarray}}

\renewcommand{\k}{\bm{k}}
\newcommand{\q}{\bm{q}}

\newcommand{\N}{\textrm{N}}
\newcommand{\Ss}{\textrm{S}}

\newcommand\m[1]{\begin{pmatrix}#1\end{pmatrix}}

\providecommand{\e}[1]{\ensuremath{{\scriptscriptstyle E\negthinspace #1}}}

\begin{document}
\allowdisplaybreaks

\author{Gy\H{o}z\H{o} Kov\'acs}
\email{kovacs.gyozo@wigner.hu}
\author{P{\'e}ter Kov\'acs}
\email{kovacs.peter@wigner.hu}
\affiliation{ELTE E{\"o}tv{\"o}s Lor\'and University, Institute of Physics, P\'azm\'any P{\'e}ter s{\'e}t\'any 1/A, 1117 Budapest, Hungary}
\affiliation{Institute for Particle and Nuclear Physics, Wigner Research Centre for Physics, Konkoly-Thege Mikl{\'o}s \'ut 29-33, 1121 Budapest, Hungary}
\author{Zsolt Sz{\'e}p}
\email{szepzs@achilles.elte.hu}
\affiliation{MTA-ELTE Theoretical Physics Research Group, P\'azm\'any P{\'e}ter s{\'e}t\'any 1/A, 1117 Budapest, Hungary}

\title{One-loop constituent quark contributions to the vector and axial-vector meson curvature mass} 

\begin{abstract}
The renormalized contribution of fermions to the curvature masses of vector and axial-vector mesons is derived with two different methods at leading order in the loop expansion applied to the $(2+1)$-flavor constituent quark-meson model. The corresponding contribution to the curvature masses of the scalar and pseudoscalar mesons, already known in the literature, is rederived in a transparent way. The temperature dependence of the curvature mass of various (axial-)vector modes obtained by decomposing the curvature mass tensor is investigated along with the \hbox{(axial-)}vector--(pseudo)scalar mixing. All fermionic corrections are expressed as simple integrals that involve at finite temperature only the Fermi-Dirac distribution function modified by the Polyakov-loop degrees of freedom. The renormalization of the \hbox{(axial-)}vector curvature mass allows us to lift a redundancy in the original Lagrangian of the globally symmetric extended linear sigma model, in which terms already generated by the covariant derivative were reincluded with different coupling constants.
\end{abstract}

\maketitle

\section{Introduction \label{sec:Intro}}

The extension of the linear sigma model with vector and axial-vector degrees of freedom has a long history (see {\it e.g.} \cite{Gasiorowicz:1969kn, Pisarski:1994yp, Denis}). In recent years, much effort was invested in the study of the phenomenological applicability of various formulations of the model. It turned out, for example, that the gauged version of the model cannot reproduce the correct decay width of the $\rho$ and $a_1$ mesons \cite{Parganlija:2008xy}, and therefore the interest shifted toward versions of the model which are based on the global chiral symmetry: originally constructed for two flavors in \cite{Parganlija:2010fz} the extended linear sigma model (EL$\sigma$M) was formulated for three flavors in \cite{Parganlija:2012fy}. 

The parametrization of the three-flavor EL$\sigma$M in relation with hadron vacuum spectroscopy was thoroughly investigated in \cite{Parganlija:2012fy}. Constituent quarks were incorporated in the EL$\sigma$M in \cite{Kovacs:2016juc} and their effect on the parametrization, through the correction induced in the curvature masses of the scalar and pseudoscalar mesons, was investigated along with the chiral phase transition at finite temperature and density. It is interesting to know how the model parameters and the results obtained in \cite{Kovacs:2016juc} are influenced by coupling the constituent quarks to the \hbox{(axial-)}vector mesons. The effect of the \hbox{(axial-)}vector mesons on the chiral transition was studied in \cite{Struber:2007bm} in the gauged version of the purely mesonic linear sigma model with chiral $U(2)_L\times U(2)_R$ symmetry, by using a rather crude approximation for the Lorentz tensor structure of the \hbox{(axial-)}vector curvature mass matrix, which was assumed to have the vacuum form even at finite temperature. Further investigations in the above-mentioned directions require the calculation of the mesonic and/or fermionic contribution to the \hbox{(axial-)}vector curvature mass matrix and its proper mode decomposition, as was done in many models dealing with the description of hot and/or dense nuclear matter \cite{Herrmann:1993za, Shiomi:1994gu, Mishra:2001xz}. Such a calculation within the linear sigma model would allow for a comparison with in-medium properties of the \hbox{(axial-)}vector mesons obtained with functional renormalization group (FRG) techniques in \cite{Eser:2015pka, Rennecke:2015eba, Jung:2016yxl, Jung:2019nnr}.

The curvature masses of the scalar and pseudoscalar mesons were derived in the $U(3)_L\times U(3)_R$ symmetric constituent quark model in \cite{Schaefer:2008hk}. The method used there involved taking the second derivative with respect to the fluctuating bosonic field of the ideal gas formula for the partition function in which the quark masses depend on these bosonic fields. The result was subsequently used in a plethora of publications, even when it does not apply, as was the case of Ref.~\cite{Tawfik:2014gga}, which seemingly uses incorrectly the result of \cite{Schaefer:2008hk} to study the effect of the temperature and chemical potential on the vector and axial-vector masses. The result derived in \cite{Schaefer:2008hk} for \hbox{(pseudo)}scalar mesons cannot be directly applied for \hbox{(axial-)}vector mesons, simply because it is not enough to consider only the boson fluctuation-dependent fermion masses: due to their Lorentz index the momentum and \hbox{(axial-)}vector fields couple to form a Lorentz scalar in the fermion determinant resulting from the fermionic functional integral. Due to such terms, derivatives of the fermionic functional determinant with respect to the \hbox{(axial-)}vector fields give additional contributions, compared to the bosonic case.

Although the calculation of the leading order fermionic contribution to the \hbox{(axial-)}vector curvature mass matrix can be done by taking the second field derivative of the functional determinant, it is much easier to take an equivalent approach and compute the self-energy at vanishing momentum with standard Feynman rules. The technical issues that need to be addressed are the mode decomposition and renormalization of the self-energy and the mixing between the \hbox{(axial-)}vector and \hbox{(pseudo)}scalar mesons.

We also mention that while our focus here is on the curvature mass, the pole mass and screening mass can also be obtained from the analytic expression of the self-energy calculated at nonzero momentum using the usual definitions given in Eq.~(6) of \cite{Helmboldt:2014iya}, where the relation between the pole and curvature masses of the mesons was investigated with FRG techniques within the two-flavor quark-meson model. This difference depends on the approximation used to solve the $O(N)$ and quark-meson models and it is typically larger for the sigma than the pion \cite{Helmboldt:2014iya, Pawlowski:2017gxj, Marko:2017yvl, Marko:2019rsl}.

The organization of the paper is as follows. In Sec.~\ref{sec:GA} an approximation scheme is presented for a consistent computation of the effective potential (pressure) in the EL$\sigma$M which is based on curvature masses that include the fermionic correction at one-loop level. In Sec.~\ref{sec:curv-mass-Nf1} we compute in the one-flavor case, $N_f=1$, the curvature mass matrix of the mesons, with both methods mentioned above. This allows for the introduction of the relevant integrals used also in Sec.~\ref{sec:SE_Nf_3}, where the self-energy of all the mesons is calculated at vanishing momentum for $N_f=2+1$ flavors. In this case a direct calculation of the curvature masses from the functional determinant, although completely straightforward, is made cumbersome by the large number of fields and the dimension of the matrix involved. This calculation is relegated to Appendix~\ref{app:brute_force}. Based on the mode decomposition of the \hbox{(axial-)}vector  self-energy, presented in detail in Appendix~\ref{app:decomp}, the curvature masses of the physical modes are given in terms of simple integrals. We also show in Sec.~\ref{sec:SE_Nf_3} how to connect the expressions of the \hbox{(pseudo)}scalar curvature masses derived here with existing ones obtained with the alternative method of Ref.~\cite{Schaefer:2008hk}. In Sec.~\ref{sec:ren} we discuss the renormalization of the \hbox{(axial-)}vector curvature masses in the isospin symmetric case. Dimensional regularization was used in order to comply with the property of the vacuum vector self-energy observed for some flavor indices, which is related to current conservation, as discussed in Appendix~\ref{app:PiV_prop}. The renormalization process revealed that the Lagrangian of the EL$\sigma$M can be written more judiciously compared to the form used in the literature, such that each term allowed by the chiral symmetry is included only once, in accordance with the generally accepted procedure. By looking from a new perspective at the wave-function renormalization factor related to the \hbox{(axial-)}vector--(pseudo)scalar mixing, we discuss in Sec.~\ref{sec:mixing} how the self-energy corrections modify its tree-level expression. Section~\ref{sec:num} contains numerical results concerning the temperature evolution of the meson masses obtained in a new vacuum parametrization of the model which takes into account the one-loop fermionic correction in the curvature mass of all the mesons. Section~\ref{sec:sum} is devoted to conclusions and an outlook. The appendixes not mentioned here contain some further technical aspects used in the calculations.

\section{Localized Gaussian approximation in the Yukawa model \label{sec:GA}}

In order to motivate our interest in the curvature mass, we present an improved calculational scheme for the effective potential of the EL$\sigma$M compared to that used in \cite{Kovacs:2016juc}. This scheme, which we call the \emph{localized Gaussian approximation}, uses the curvature mass of the various mesons. To keep the notation simple, we consider the simplest chirally symmetric Yukawa model, defined by the Lagrangian 
\be
{\cal L}_{\rm Y}=\frac{1}{2} \partial_\mu\varphi \partial^\mu\varphi - U_{\rm cl}(\varphi)
+ \bar\psi\big(i \slashed{\partial} - g_S\varphi)\psi,
\label{Eq:L_Y}
\ee
where $\psi$ and $\varphi$ are fermionic and bosonic fields and $U_\textrm{cl}(\varphi)=m^2_0\varphi^2/2+\lambda\varphi^4/24$ is the classical potential. We use Minkowski metric $g^{\mu\nu}=\textrm{diag}(1,-1,-1,-1)$ and the conventions of Ref.~\cite{PS}. 

Integrating over the fermions in the partition function\footnote{In the L$\sigma$M this step is motivated by the fact that $\psi$ represents the constituent quark, that is an effective degree of freedom not necessarily corrected by the interaction with mesons.}
\be
Z = \int{\cal D}\varphi {\cal D}\bar\psi{\cal D}\psi e^{i \int_x {\cal L}_{\rm Y}}
= \int{\cal D}\varphi  e^{i {\cal A}(\varphi) },
\ee
leads to the action ($\int_x\equiv d^4x$)
\be
{\cal A}(\varphi) =\int_x
\left[\frac{1}{2} \partial_\mu\varphi \partial^\mu\varphi - U_{\rm cl}(\varphi)\right]
-i \Tr\log\big(i \mathcal{S}^{-1}(\slashed{\partial};\varphi)\big),
\label{Eq:eff_action}
\ee
where $\Tr\equiv \tr_{\rm D}\int d^4x$ denotes the functional trace, with the subscript ``D'' referring to the Dirac space, and
\be
i\mathcal{S}^{-1}(x,y)=\big[i \slashed\partial_x - g_S\varphi(x)\big]\delta^{(4)}(x-y)
\ee
is the inverse fermion propagator.

Shifting the field with an $x$-independent background $\phi$, $\varphi(x)\to\phi+\varphi(x)$, the effective potential can be constructed along the lines of Ref.~\cite{Jackiw:1974cv}. Several approximations of the effective potential are considered in the literature.

\paragraph{Mean-field approximation} The bosonic fluctuating field is neglected altogether, leading to
\be
U_{\rm MF}(\phi)=U_{\rm cl}(\phi) + i \tr_{\rm D}\int_K\log\big(i \mathcal{S}_f^{-1}(K)\big),
\label{Eq:Veff_MF}
\ee
where $i\mathcal{S}_f^{-1}(K)=\slashed{K}-m_f$ is the tree-level fermion inverse propagator with mass $m_f=g_S\phi$. Here we introduced the notation $\int_K \equiv\int \frac{d^4 K}{(2\pi)^4}$ for the momentum integral with 4-momentum $K^\mu=(k_0,\k).$ The field equation used in \cite{Kovacs:2016juc} was derived in this approximation.

\paragraph{Ideal gas approximation} The bosonic fluctuating field is neglected in the fermion determinant ($\Tr\log$) appearing in Eq.~\eqref{Eq:eff_action} and kept only to quadratic order in the terms coming from the expansion of $U_{\rm cl}(\phi+\varphi)$. The Gaussian functional integral over $\varphi$ leads to
\be
U_{\rm eff}^{\rm IG}(\phi)=U_{\rm MF}(\phi) -\frac{i}{2}\int_K\log\big( i{\cal D}^{-1}(K;\phi)\big),
\label{Eq:Veff_IG}
\ee
where $i{\cal D}^{-1}(K;\phi)=K^2-\hat m^2(\phi)$ is the tree-level boson propagator with $\hat m^2(\phi)=d^2 U_{\rm cl}(\phi)/d\phi^2$ being the classical curvature mass. This approximation was used in a nonsystematic way in \cite{Kovacs:2016juc} to include mesonic corrections in the pressure.

\paragraph{Ring resummation or Gaussian approximation} The fermion determinant is expanded in powers of $\varphi$ and keeping in Eq.~\eqref{Eq:eff_action} the term  quadratic in the fluctuating mesonic field, the Gaussian functional integral over $\varphi$ results in
\be
U_{\rm eff}^{\rm GA}(\phi)=U_{\rm MF}(\phi) -\frac{i}{2}\int_K\log\big( i{\cal D}^{-1}(K;\phi)-\Pi(K;\phi)\big),
\label{Eq:Veff_GA}
\ee
where the boson self-energy 
\be
\Pi(K;\phi) = i g_S^2 \tr \int_P \mathcal{S}_f(P) \mathcal{S}_f(K-P),
\label{Eq:Pi_simple}
\ee
represents the one-loop contribution of the fermions. Expanding in Eq.~\eqref{Eq:Veff_GA} the logarithm one recognizes the integrals of the ring resummation.

The ring resummation is widely used in the Nambu--Jona-Lasinio model, where it goes by the name of random-phase approximation \cite{Klevansky:1992qe}. In that context the integral in Eq.~\eqref{Eq:Veff_GA} requires no renormalization and was evaluated using cutoff regularization in \cite{Yamazaki:2012ux, Torres-Rincon:2017zbr}. To spare the trouble of renormalizing this integral in a linear sigma model, one can approximate the self-energy with its zero momentum limit. In this \emph{localized} approximation the dressed bosonic inverse propagator appearing in Eq.~\eqref{Eq:Veff_GA} is of tree-level type, just that the tree-level mass is replaced by the one-loop curvature mass $\hat M^2(\phi) \equiv \hat m^2(\phi) + \Pi(K=0;\phi)$. Since with a homogeneous scalar background the curvature mass does not depend on the momentum, the renormalization of the integral becomes an easy task, as discussed in \cite{Marko:2015gpa} (see also Eq.~\eqref{Eq:tr-log_int_nonconv_ren} in Sec.~\ref{sec:ren}). 

Note that one can define a curvature mass in each of the above approximations, by taking the second derivative of the potential in Eq.~\eqref{Eq:Veff_MF}, \eqref{Eq:Veff_IG}, or \eqref{Eq:Veff_GA} with respect to the field $\phi.$ The curvature mass we investigate in this paper contains the fermionic contribution from the second field derivative of the $\Tr\log$ in Eq.~\eqref{Eq:eff_action}. This represents the purely fermionic one-loop contribution to the curvature mass which can be derived in principle in the localized Gaussian approximation using the background field method.

In order to compute the pressure, we need to evaluate the effective potential at the minimum. In the localized approximation the extremum, is determined as the solution of the field equation
\bea
&&m_0^2\phi+\frac{\lambda}{6}\phi^3+\frac{1}{2}\left[\lambda\phi + 2 g_S^3 \tr_{\rm D}\int_P \mathcal{S}_f^3(P) \right]\nonumber\\
&&\times\int_K\frac{i}{K^2-\hat M^2(\phi)} - g_S\tr_{\rm D}\int_K\mathcal{S}_f(K)=0.
\eea
We mention that the second term in the square brackets is nothing but the fermionic correction to the three-point vertex function evaluated at vanishing momentum. This vertex function is obtained by expanding the fermionic determinant in powers of the bosonic field
\bea
\Tr\log \big(i {\cal S}_f^{-1}-g\varphi\big) = \Tr\log \big(i {\cal S}_f^{-1}\big) - \sum_{n=1}^\infty \frac{(-ig)^n}{n}\nonumber \\
\times\tr_{\rm D}\bigg[\prod_{i=1}^n\int {\rm d}^4 x_i\, \varphi(x_i)  {\cal S}_f(x_i,x_{i+1})\bigg]_{x_{n+1}=x_1}. \qquad
\eea
The expansion gives a series of one-loop diagrams in which the $n$th term has $n$ external fields (see {\it e.g.} \cite{Eguchi:1976iz} or Ch.~9.5 of \cite{PS}). Using the background field method, the expansion of such a fermionic functional determinant was considered recently in \cite{Braghin:2018vyr, Braghin:2020enw} in order to derive effective couplings between constituent quarks, (axial-)vector mesons, and the photon.

The second field derivative of the functional determinant, taken at vanishing mesonic field, is nothing but the one-loop self-energy associated with the bosonic field with respect to which the derivative is taken, as the contribution of diagrams not having exactly two external fields of this type vanishes. Based on this observation one can obtain the lowest order fermionic correction to the bosonic curvature mass by computing the one-loop self-energy with standard Feynman rules.

\section{Curvature mass in the $N_f=1$ case \label{sec:curv-mass-Nf1}}

We generalize now Eq.~\eqref{Eq:L_Y} and consider the chirally symmetric Lagrangian\footnote{The one-loop curvature mass formulas derived here can be easily modified when, in the absence of chiral symmetry, $P$ and $A$ can have different Yukawa couplings than $S$ and $V$, respectively.} in which a fermionic field $\psi$ interacts through a Yukawa term to scalar ($S$), pseudoscalar ($P$), vector ($V_\mu$), and axial-vector ($A_\mu$) fields
\be
   {\cal L}_f=\bar\psi\left[i\gamma^\mu\partial_\mu-g_S(S-i\gamma_5 P) -g_V \gamma^\mu \big(V_\mu + \gamma_5 A_\mu\big) \right]\psi.
\label{Eq:L_f}
\ee

The mesonic part of the Lagrangian is of the form
\bea
\nonumber
&&{\cal L}_\textrm{m}=\sum_{X=S,P}\Big[\frac{(\partial X)^2}{2}-\frac{m^2_0}{2} X^2\Big]-\frac{\lambda}{4!}\big(S^2+P^2\big)^2\\
&&+\sum_{Y=V,A}\Big[-\frac{1}{4}F_{\mu\nu}^YF_Y^{\mu\nu}+\frac{m^2_{0,V}}{2}Y^\mu Y_\mu\Big] + {\cal L}_{\rm m}^\textrm{int}(X,Y_\mu),\quad
\label{Eq:L_mesonic}    
\eea
with $F_{\mu\nu}^Y=\partial_\mu Y_\nu - \partial_\nu Y_\mu.$ We shall return to the unspecified interacting part in the $N_f=2+1$ case in relation to the renormalization of the one-loop curvature masses.

By integrating over the fermions in the partition function, done after the usual shifts $S(x)\to \phi+S(x)$ ($\phi$ is independent of $x$) introduced in order to deal with the spontaneous symmetry breaking (SSB), one obtains a correction to the classical mesonic action in the form of a functional determinant. The expansion of the functional determinant in powers of mesonic field derivatives, the so-called derivative expansion \cite{Aitchison:1985pp, Chan:1985ny}, leads to an effective bosonic action of the form
\be
\int d^4x {\cal L}_\textrm{eff}(\phi,\xi)=\int d^4x \big[{\cal L}_\textrm{m}(\phi,\xi)-U_f(\phi,\xi) + {\cal O}((\partial \xi)^2)\big],
\ee
with the leading order term of the expansion being the one-loop fermionic effective potential 
\bea
U_f(\phi,\xi(x)) &=& i \Tr \log\big(i \mathcal{S}^{-1}_f(\slashed\partial;\xi)\big)\nonumber \\
&=&i \int_K \log {\rm det}_{\rm D}\big(i \mathcal{S}_f^{-1}(K;\xi(x))\big),
\label{Eq:U_f}
\eea
which depends on all the fluctuating mesonic fields collectively denoted by $\xi(x)$. We introduced
\be
i \mathcal{S}_f^{-1}(K;\xi)=\slashed K-m_f-g_S\big(S -i\gamma_5P\big)-g_V\big(\slashed V+\slashed A \gamma_5),
\ee
for the inverse fermion propagator, in which $m_f=g_S \phi$ is the tree-level (classical) fermion mass. Hereafter the $x$ dependence of the mesonic fields will not be indicated.

The second derivative of $U_f(\phi,\xi)$ with respect to the mesonic fields gives an additive correction to the classical mesonic curvature mass obtained from ${\cal L}_\textrm{m}$. Since later we will investigate the $N_f=2+1$ case, where the fields have flavor indices $a=0,\dots 8$, we give the more general formulas of these corrections
\be
\begin{aligned}
\Delta \hat m^{2,(X)}_{ab}&\equiv\frac{d^2 U_f(\phi,\xi)}{d X_a d X_b} \bigg|_{\xi=0}, \quad & X=S,P,\\
\Delta \hat m^{2,(Y)}_{ab,\mu\nu}&\equiv-\frac{d^2 U_f(\phi,\xi)}{d Y^\mu_a d Y^\nu_b} \bigg|_{\xi=0}, \quad & Y=V,A.
\end{aligned}
\label{Eq:dhatM2_SPVA}
\ee
In this section the flavor indices should be disregarded.

The sign difference between the above definitions is due to the different signs of the corresponding classical mass terms in Eq.~\eqref{Eq:L_mesonic}. Accordingly, for the \hbox{(pseudo)}scalar field one has an additive correction to the classical mass squared $\hat m^{2,(X)}_{ab}\equiv -\frac{d^2{\cal L}_\textrm{m}}{d X_a d X_b}\big|_{\xi=0},$ while for the \hbox{(axial-)}vector field the second derivative is a Lorentz tensor, and therefore the correction to $\hat m^{2,(Y)}_{ab}\equiv \frac{g^{\nu\mu}}{4} \frac{d^2{\cal L}_\textrm{m}}{d Y^\mu_a d Y^\nu_b}\big|_{\xi=0}$ depends on the tensor structure of $\Delta \hat m^{2,(Y)}_{ab,\mu\nu}$ and whether one works at zero or nonzero temperature. For example, at $T=0$, where $\Delta \hat m^{2,(Y)}_{ab,\mu\nu}\propto g_{\mu\nu}$, the fermionic correction to $\hat m^{2,(Y)}_{ab}$ is obtained by taking the trace in Eq.~\eqref{Eq:dhatM2_SPVA}:
\be
\Delta \hat m^{2,(Y)}_{ab}:= \frac{1}{4}\Delta\hat m^{2,(Y)\mu}_{ab,\mu}.
\label{Eq:dhatm2_VA}
\ee
This is needed in a parametrization of the model that is based on the one-loop curvature masses. For the curvature mass at $T\ne0$ one needs the mode decomposition of the tensor $\Delta \hat m^{2,(Y)}_{\mu\nu}.$ This will be discussed in Sec.~\ref{sec:SE_Nf_3} and Appendix~\ref{app:decomp} in relation to the $N_f=2+1$ case.

\subsection{Brute force calculation \label{ss:brute_1}}

The determinant $D\equiv\det_{\rm D}\big(i \mathcal{S}_f^{-1}(K;\xi)\big)$ appearing in Eq.~\eqref{Eq:U_f} evaluates to
\begin{align}
D &= \big[g_S^2\big((S+\phi)^2+P^2\big)-K^2\big]^2\nonumber\\
&+ \big(K^2-m_f^2+g_V^2 V^2 - 2 g_V K\cdot V\big)^2\nonumber\\
&+ \left[\big(g_V^2 A^2\big)^2 + 2 g_V^2 \big[(m_f^2+K^2) A^2 - 2(K\cdot A)^2\big]\right]\nonumber\\
&+2 g_S^2\big[(S+\phi)^2+P^2-\phi^2\big]\big[g_V^2 \big(A^2 - V^2\big) + 2 g_V K\cdot V \big]\nonumber\\
&-\big(K^2-m_f^2\big)^2.
\label{Eq:det_Nf_1}
\end{align}

Writing the determinant in this form facilitates the derivation of the curvature masses, as the contribution to the scalar and the pseudoscalar comes only from the first term, while only the second and the third terms contribute in the case of the vector and the axial-vector, respectively. Also, note that $D(\xi=0)=\big(K^2-m_f^2\big)^2.$

The second derivative of the determinant with respect to a particular field denoted by $\varphi$ is calculated using
\be
\frac{d^2 \log D}{d\varphi d\varphi}\bigg|_{\xi=0}=\left[
  \frac{1}{D} \frac{d^2 D}{d\varphi d\varphi} - \frac{1}{D^2}\frac{d D}{d\varphi}\frac{d D}{d\varphi}
  \right]\bigg|_{\xi=0}.
\label{Eq:diff_log_D}
\ee
For $\varphi\in\{S,P,V_\mu\}$ this is applied writing $D=\tilde D^2 +R,$ where $\tilde D^2$ is either the first or the second term on the right-hand side (rhs) of Eq.~\eqref{Eq:det_Nf_1}, while the remnant $R$ does not contribute in Eq.~\eqref{Eq:diff_log_D}. Introducing the notation $G_f(K)=1/(K^2-m_f^2)$ one obtains
\begin{subequations}
\bea
\frac{d^2 \log D}{dS dS}\bigg|_{\xi=0}&=&-4 g_S^2\left[G_f+2m_f^2 G_f^2\right],\\
\frac{d^2 \log D}{dP dP}\bigg|_{\xi=0}&=&-4 g_S^2 G_f,
\eea
\label{Eq:diff_log_D_SP}
\end{subequations}
\hspace*{-0.19cm}for the scalar and the pseudoscalar fields, and
\begin{subequations}
\be
\frac{d^2 \log D}{d V^\mu d V^\nu}\bigg|_{\xi=0}=4g_V^2\big[g_{\mu\nu} G_f - 2 K_\mu K_\nu G_f^2\big]\,,
\label{Eq:diff_log_D_V}
\ee 
for the vector field.

For the axial-vector one applies Eq.~\eqref{Eq:diff_log_D} with $D=\tilde D + R$, where $\tilde D$ is the third term on the rhs of Eq.~\eqref {Eq:det_Nf_1}, to obtain
\bea
\frac{d^2 \log D}{d A^\mu d A^\nu}\bigg|_{\xi=0}=4 g_V^2\big[(m_f^2+K^2)g_{\mu\nu} - 2 K_\mu K_\nu\big]G_f^2.\qquad
\label{Eq:diff_log_D_A}
\eea
\label{Eq:diff_log_D_VA}
\end{subequations}
\hspace*{-0.125cm}For scalar and vector fields there are contributions from both the first and the second derivative of $\tilde D$, while in the case of the pseudoscalar and axial-vector fields only the second derivative of $\tilde D$ contributes.

Using Eqs.~\eqref{Eq:U_f}, \eqref{Eq:dhatM2_SPVA}, \eqref{Eq:det_Nf_1}, and \eqref{Eq:diff_log_D_SP} the fermion corrections to the curvature masses of the scalar and pseudoscalar fields are obtained as
\begin{subequations}
\bea
\Delta \hat m^{2,(S)}&=&-4g_S^2\bigg[1 + 2 m_f^2\frac{d}{d m_f^2} \bigg]{\cal T}(m_f),\\
\Delta \hat m^{2,(P)}&=&-4g_S^2{\cal T}(m_f),
\eea
\label{Eq:dhatm2_SP_final}
\end{subequations}
\hspace*{-0.125cm}where the (\emph{vacuum}) tadpole integral is  
\be
{\cal T}(m_f) = \int_K \frac{i}{K^2-m_f^2} = \int_K i G_f(K).
\label{Eq:tad}      
\ee

In the case of the vector and the axial-vector fields, one evaluates the trace in Eq.~\eqref{Eq:dhatm2_VA}, using Eqs.~\eqref{Eq:dhatM2_SPVA} and \eqref{Eq:diff_log_D_VA} together with $g_\mu^{\ \mu}=4$ and $K_\mu K_\nu g^{\nu\mu}=K^2=m_f^2+G_f^{-1}$, to obtain 
\begin{subequations}
\bea
\Delta \hat m^{2,(V)}&=&-2g_V^2\bigg[1 - m_f^2\frac{d}{d m_f^2} \bigg]{\cal T}(m_f), \label{Eq:dhatm2_V_final}\\
\Delta \hat m^{2,(A)}&=&-2g_V^2\bigg[1 + 3 m_f^2\frac{d}{d m_f^2} \bigg]{\cal T}(m_f).
\eea
\label{Eq:dhatm2_VA_final}
\end{subequations}

\subsubsection{Integrals at finite temperature}

The expressions in Eqs.~\eqref{Eq:dhatm2_SP_final} and \eqref{Eq:dhatm2_V_final}, which were formally derived at vanishing temperature ($T=0$), can be easily generalized to $T\ne0$, where the tadpole integral consists of \emph{vacuum} and \emph{matter} parts:
\be
{\cal T}(m_f) = {\cal T}^{(0)}(m_f) + {\cal T}^{(1)}(m_f).
\ee
The superscripts indicate the absence or the presence of statistical factors in the respective integrands. In a covariant calculation the \emph{vacuum} part $T^{(0)}(m_f)$ is the integral in Eq.~\eqref{Eq:tad}, while in a noncovariant calculation it is
\be
T^{(0)}(m_f)=\int\frac{d^3 k}{(2\pi)^3}\frac{1}{2 (\k^2 +m_f^2)^{1/2}},
\label{Eq:tad_vac_noncov}
\ee
as obtained with the usual conventions of the imaginary time formalism \cite{Kapusta-Gale}, namely ($\mu$ is the chemical potential)
\be
k_0\to i\nu_n+\mu \quad \textrm{and} \quad \int_K\to i T\underset{n}{\sum}\int \frac{d^3 k}{(2\pi)^3},
\label{Eq:ITF}
\ee
after doing the summation over the fermionic Matsubara frequencies $\nu_n=(2 n + 1) \pi T$. The \emph{matter} part is
\be
{\cal T}^{(1)}(m_f)=-\frac{1}{4\pi^2}\int_0^\infty\textrm{d} k\frac{k^2}{E_f(k)}\big[f_f^+(k)+f_f^-(k)\big],\label{Eq:tad_FD_T}
\ee
where $f_f^\pm(k)=1/(\exp((E_f(k)\mp \mu)/T)+1)$ are the Fermi-Dirac distribution functions for particles and antiparticles and $E_f^2(k)=k^2+m_f^2$ with $k=|\k|$.

For the mass derivative of the \emph{matter} part of the tadpole integral (Euclidean bubble integral at vanishing momentum)  one uses  $d(f_f^\pm/E_f)/d m_f^2 = d(f_f^\pm/E_f)/d k^2$ and an integration by parts to obtain
\be
-{\cal B}^{(1)}(m_f)\equiv \frac{d{\cal T}^{(1)}(m_f)}{d m_f^2}=\frac{1}{8\pi^2}\int_0^\infty d k\frac{f_f^+(k)+f_f^-(k)}{E_f(k)}. \label{Eq:Bub_T}
\ee

The fact that even at finite temperature the trace of the second derivative appearing in Eq.~\eqref{Eq:diff_log_D_V} is the only relevant quantity determining the curvature mass of the vector boson is due to current conservation. For the axial-vector this is not the case and one needs the mode decomposition of the tensor in Eq.~\eqref{Eq:diff_log_D_A}. This is discussed in Appendix~\ref{app:decomp}.

\subsection{Curvature mass from the self-energy \label{ss:SE_Nf_1}}

As mentioned at the end of Sec.~\ref{sec:GA}, the one-loop curvature mass can also be obtained by computing the corresponding zero momentum self-energy. For example, for the self-energy of the vector field, the Feynman rules applied with the conventions of \cite{PS} give
\be
i\Pi_{\mu\nu}^{(V)}(Q=0)= -\big(-ig_V\big)^2\tr_{\rm D}\int_K\gamma_\mu {\cal S}_f(K)\gamma_\nu{\cal S}_f(K).
\label{Eq:polarisation_def}
\ee
Using $S_f=i(\slashed K + m_f) G_f$, the Dirac trace 
\be
\tr_{\rm D}\big[\gamma_\mu(\slashed K +m_f)\gamma_\nu(\slashed K +m_f)\big]= 8 K_\mu K_\nu - 4 g_{\mu\nu}G_f^{-1}(K),
\label{Eq:vector-tr}
\ee
results in
\be
\Pi_{\mu\nu}^{(V)}(Q=0) = -4 g_V^2 \bigg[ g_{\mu\nu} {\cal T}(m_f)-2i\int_K K_\mu K_\nu G_f^2(K)\bigg].
\label{Eq:polarisation_int}
\ee
Comparing Eq.~\eqref{Eq:polarisation_int} with the expression obtained by using  Eq.~\eqref{Eq:diff_log_D_V} in Eq.~\eqref{Eq:dhatM2_SPVA} shows explicitly that $\Delta \hat m^{2,(V)}_{\mu\nu}=\Pi_{\mu\nu}^{(V)}(Q=0)$. At zero temperature $\Pi_{\mu\nu}^{(V)}(0)=0$ due to the current conservation related to the $U(1)_V$ global symmetry, and therefore the one-loop curvature mass of the vector boson is the classical one, $\hat m^{2,(V)}$.

In order to obtain the curvature mass of the physical modes at finite temperature, we need the standard decomposition of the momentum-dependent self-energy tensor reviewed in Appendix~\ref{app:decomp}. The self-energy is decomposed into \emph{vacuum} and \emph{matter} parts. The former is evaluated using a covariant calculation performed at $T=0$ in a regularization scheme compatible with the consequence of current conservation, namely that the self-energy is 4-transverse, $Q^\mu\Pi_{\mu\nu}^{(V)}(Q)=0,$ and $\Pi^{(V)}_{\mu\nu,\textrm{vac}}(Q=0)=0.$ Therefore, only the \emph{matter} part contributes to the self-energy components determining the curvature masses of the 3-longitudinal and 3-transverse vector modes
\be
\hat M^{2,(V)}_\textrm{l/t} = \hat m^{2,(V)}+ \Pi^{(V)}_\textrm{l/t}.
\ee
The components are obtained as (see Ch.~1.8 of \cite{Das})
\be
\begin{aligned}
    \Pi^{(V)}_\textrm{l}&=-\lim_{\q\to 0}\lim_{q_0\to 0} \frac{Q^2}{\q^2} \Pi^{(V),00}(Q) = \Pi^{(V),00}_\textrm{mat}(0,{\bf 0}),\\
    \Pi^{(V)}_\textrm{t}&=\frac{1}{2}\lim_{\q\to 0}\lim_{q_0\to 0}\big(\Pi^{(V),\mu}_\mu(Q)-\Pi_\textrm{l}(Q)\big)\\
    &= -\frac{3}{2}\Pi^{(V),11}_\textrm{mat}(0,{\bf 0}).
\end{aligned}
\ee
For the axial-vector, which does not couple to a conserved current, the tensor structure of the self-energy is more complicated and it is discussed in Appendix~\ref{app:decomp}.

The interested reader can find in Appendix~\ref{app:check_comp} a discussion on the \emph{matter} part of the self-energy components. For the \emph{vacuum} part see the discussion in Sec.~\ref{sec:ren} and the calculation presented in Appendix~\ref{app:PiCalcDetails}.

\section{Curvature mass for $N_f=2+1$ \label{sec:SE_Nf_3}}

The fermionic part of the chiral-invariant Lagrangian of the extended linear sigma model, whose mesonic part can be found in \cite{Kovacs:2016juc}, has the form given in Eq.~\eqref{Eq:L_f}, only that the fermionic field is the triplet of constituent quarks, $\psi\equiv(u,d,s)^{\rm T}$, while the mesonic fields are nonets. For the scalars $S=S_a T_a = S_a \lambda_a/2$, $a=0,\dots,8$ and similarly for the other mesons ($\lambda_{a\ne 0}$ are the Gell-Mann matrices and $\lambda_0=\sqrt{\frac{2}{3}}\mathbb{1}$).

The integration over the fermionic field in the partition function results in a functional determinant involving a $N\times N$ matrix, where $N=3\times 4 \times N_c$ with $N_c$ being the number of colors. This matrix structure makes tedious a brute force calculation of the curvature mass similar to the one shown in Sec.~\ref{ss:brute_1}, even in the case of a trivial color dependence  (see Eq.~\eqref{Eq:matrix_Nf3}). Therefore, we proceed as in Sec.~\ref{ss:SE_Nf_1} by calculating the self-energy at vanishing momentum and relegate to Appendix~\ref{app:brute_force} the sketch of a direct calculation.

\subsection{Curvature mass from the self-energy \label{ss:SE_Nf_3}}

For simplicity, we consider the case when only the scalar fields, namely $S_0, S_3$ and $S_8,$ have expectation values, denoted by $\phi_0,\phi_3,$ and $\phi_8.$ It proves convenient to work in the $\N-\Ss$ (nonstrange--strange) basis, which for a generic quantity $Q$ is related to the $(0,8)$ basis by
\be
\m{Q_\N\\Q_\Ss} = R \m{Q_0\\Q_8},\  R=R^{-1}=\frac{1}{\sqrt{3}}\m{\sqrt{2} & 1\\ 1 & -\sqrt{2}}.
\ee
Applying the above relation to the matrices $\lambda_0$ and $\lambda_8$, one obtains $\lambda_\N=\textrm{diag}(1,1,0)$ and $\lambda_\Ss=\sqrt{2}\,\textrm{diag}(0,0,1)$, which give the antisymmetric structure constants $f_{45{\rm N}}=f_{67{\rm N}}=1/2$ and $f_{45{\rm S}}=f_{67{\rm S}}=-1/\sqrt{2}.$ With the shift $S_i\to \phi_i+S_i, i=\N,3,\Ss,$ one obtains, using also $\lambda_3=\textrm{diag}(1,-1,0)$, the tree-level inverse fermion propagator matrix in flavor space as $i\bar{\cal S}_0^{-1}=\textrm{diag}(i{\cal S}_u^{-1},i{\cal S}_d^{-1},i{\cal S}_s^{-1}),$ with components $i{\cal S}_f^{-1}(K)=\slashed{K}-m_f,$ where the tree-level fermion masses are 
\be
m_{u,d}=g_S(\phi_\N \pm \phi_3)/2, \quad  m_s = g_S \phi_\Ss/\sqrt{2}.
\label{Eq:m_f_tree}
\ee

The one-loop self-energy of a generic field $X_a$, with $a$ being a flavor index, can be written as
\be
\Pi_{ab}^{(X)}(Q) = i N_c s_X c_X^2\int_K\tr\Big[\Gamma_X\frac{\lambda_a}{2} \bar{\cal S}_0(K)\Gamma'_X\frac{\lambda_b}{2}\bar{\cal S}_0(L)\Big],
\label{Eq:SE_Nf3}
\ee
where $N_c$ is the number of colors, $L=K-Q$, $s_X=1$ for $X=V,A$ and $s_X=-1$ for $X=S,P.$ The propagator matrix $\bar{\cal S}_0=\textrm{diag}({\cal S}_u,{\cal S}_d,{\cal S}_s)$ has as elements the tree-level propagators of the constituent quarks. Furthermore, $\Gamma_X$ contains Dirac matrices that carries a Lorentz index when $X\in\{V,A\}$, in which case the prime on $\Gamma'_X$ indicates that its Lorentz index is different from that of $\Gamma_X$. The matrices are explicitly given in Table~\ref{Tab:couplings}, along with the constant $c_X$ proportional to the Yukawa coupling. The trace is to be taken in Dirac and flavor spaces. We assumed a trivial color dependence and we postpone to Sec.~\ref{ss:prev-calc} the discussion of a more complicated one.

\begin{table}
\caption{Dirac matrices and couplings to be used in the self-energy formula \eqref{Eq:SE_Nf3}. \label{Tab:couplings} }  
\begin{tabular}{ c|| c| c| c| c }
\hline\\[-0.75em]    
$X, c_X, \Gamma_X$ & $S, -i g_S,\mathbb{1}$ & $P, -g_S,\gamma_5$ & $V, -i g_V,\gamma_\mu$ & $A, -i g_V,\gamma_\mu\gamma_5$
\\[0.2em] \hline
\end{tabular}
\end{table}

The flavor space trace in Eq.~\eqref{Eq:SE_Nf3} can be easily performed. Since in the $\N-\Ss$ basis the $\lambda_a$ matrices have, with the exceptions of $a=\Ss$, two nonzero matrix elements, one generally obtains two terms which can cancel each other for some flavor combinations. The nonzero contributions are listed in Table~\ref{Tab:flavor_tr}. In the case of the first three entries, the factor of 2 is the consequence of the identity
\bea
&&\tr_{\rm D}\big[ \Gamma_X{\cal S}_f(K) \Gamma'_Y {\cal S}_{f'}(R)\big]\nonumber\\
&&\qquad = r_X r_Y\tr_{\rm D}\big[\Gamma_X{\cal S}_{f'}(-R) \Gamma'_Y {\cal S}_{f}(-K)\big],
\eea
which is applied inside the integral in Eq.~\eqref{Eq:SE_Nf3} with $Y=X$, followed by the shift $K\to-K$. This identity can be proven using the cyclicity of the trace and that, given the charge conjugation operator $C=i\gamma^2\gamma^0,$ the matrices $\Gamma_X$ of Table~\ref{Tab:couplings} satisfy $C \Gamma_X C^{-1}=r_X\Gamma_X^\textrm{T},$ where $r_X=1$ for $X\in\{S,P,A\}$ and $r_X=-1$ for $X=V.$

\begin{table}[!t]
\caption{Nonvanishing contributions to the self-energy \eqref{Eq:SE_Nf3} from the flavor space trace, $\tr\big[\lambda_a\bar{\cal S}_0\lambda_b\bar{\cal S}_0\big],$ for $\phi_3\ne0$ and their reduction in the isospin symmetric case, where $l$ denotes the light quarks with equal masses $m_l\equiv m_u=m_d.$ \label{Tab:flavor_tr}}
  \begin{center}
  \begin{tabular}{c| c c}
\hline\hline   
$ab$ & $\phi_3\ne 0$ & $\phi_3=0$ \\
\hline
11,22 & $2{\cal S}_u {\cal S}_d$ & $2{\cal S}_l {\cal S}_l$ \\
44,55 & $2{\cal S}_u {\cal S}_s$ & $2{\cal S}_l {\cal S}_s$ \\
66,77 & $2{\cal S}_d {\cal S}_s$ & $2{\cal S}_l {\cal S}_s$ \\
SS & $2{\cal S}_s {\cal S}_s$ & $2{\cal S}_s {\cal S}_s$ \\
33,NN & $\ {\cal S}_u {\cal S}_u + {\cal S}_d {\cal S}_d\ $ & $2{\cal S}_l {\cal S}_l$ \\
3N, N3 & $\ {\cal S}_u{\cal S}_u - {\cal S}_d {\cal S}_d\ $ & $0$\\
\hline\hline
\end{tabular}
\end{center}
\end{table}

\begin{table*}
\caption{Fermionic contribution to the zero momentum one-loop self-energy of the scalar ($S$), pseudoscalar ($P$), vector ($V$) and axial-vector ($A$) fields in the $\phi_3\ne0$ case. We indicate by $f$ and $f'$ the quark type whose mass has to be taken into account in the formula of the self-energy having flavor indices $ab.$ The \emph{matter} part of the tadpole integral ${\cal T}(m)$ is given in Eq.~\eqref{Eq:tad_FD_T} and the finite piece of the \emph{vacuum} part in Eq.~\eqref{Eq:Tad_vac_F}. The \emph{vacuum} integral $I^{V/A}_{\rm vac}$ is given in Appendix~\ref{app:PiCalcDetails}, together with the $00$ and $11$ components of the \emph{matter} integral $I^{V/A,\mu\nu}_{\rm mat}$. The constants are $C_{S/V}=2N_cg_{S/V}^2$, $t_S=1,$ $t_P=0,$ and $s_{u/d}=\pm 1.$ \label{Tab:Pi_all}}  
\begin{center} 
  \begin{tabular}{ c| c c c}
    \hline\hline \\[-0.85em]    
    ab &\ \ f\ f' \ & $-\Pi^{(S/P)}_{ab}/C_S$ & $\Pi^{(V/A),\mu\nu}_{ab}/C_V$ \\[2pt] \hline\\[-0.85em]
    11, 22 & d\ u &  \quad \multirow{3}{*}{\raisebox{-0.5cm}{$\displaystyle \frac{m_f {\cal T}(m_f) \mp m_{f'} {\cal T}(m_{f'})}{m_f \mp m_{f'}}$}} & \quad \multirow{3}{*}{\raisebox{-0.4cm}{$ \displaystyle g^{\mu\nu} I^{V/A}_{\rm vac}(m_f,m_{f'}) + I^{V/A,\mu\nu}_{\rm mat}(m_f,m_{f'})$}} \\[3pt]
    44, 55 & s\ u & \\[3pt]
    66, 77 & d\ s & \\[3pt]
    SS & s\ - & \quad $\displaystyle \bigg(1 + 2t_{S/P} m_f^2 \frac{d}{d m_f^2}\bigg){\cal T}(m_f)$ & \quad $\displaystyle g^{\mu\nu} I^{V/A}_{\rm vac}(m_f) + I^{V/A,\mu\nu}_{\rm mat}(m_f) $ \\
    33, NN& u\ d & \quad $\displaystyle \frac{1}{2}\sum_{i=f,f'}\left(1+ 2 t_{S/P} m_i^2 \frac{d}{d m_i^2}\right){\cal T}(m_i)$ & \quad  $\displaystyle \frac{1}{2}\sum_{i=f,f'}\left(g^{\mu\nu} I^{V/A}_{\rm vac}(m_i) + I^{V/A,\mu\nu}_{\rm mat}(m_i)\right)$ \\
    3N, N3 \ & u\ d & \quad $\displaystyle \frac{1}{2}\sum_{i=f,f'}s_i\bigg(1 + 2 t_{S/P} m_i^2 \frac{d}{d m_i^2}\bigg){\cal T}(m_i)$ & \quad $\displaystyle \frac{1}{2}\sum_{i=f,f'}s_i\left(g^{\mu\nu} I^{V/A}_{\rm vac}(m_i) + I^{V/A,\mu\nu}_{\rm mat}(m_i)\right)$ \\[10pt]
    \hline\hline
  \end{tabular}
\end{center}
\end{table*}

We see from Table~\ref{Tab:flavor_tr} that after the trace in flavor space is performed, depending on the indices $ab$, the self-energy \eqref{Eq:SE_Nf3} can be expressed either in terms of integrals involving two different propagators 
\be
I^X(Q;m_f,m_{f'})=\frac{-i}{4}\int_K \tr_{\rm D}\big[\Gamma_X {\cal S}_f(K) \Gamma'_X {\cal S}_{f'}(K-Q)\big],
\label{Eq:I_X_def}
\ee
or using integrals of the types already encountered in the one-flavor case (see Eq.~\eqref{Eq:polarisation_def}), obtained from Eq.~\eqref{Eq:I_X_def} as
\be
I^X(Q;m_f)=\lim_{m_{f'}\to m_f}I^X(Q;m_f,m_{f'}).
\label{Eq:I_X_one_mass}
\ee

Being interested in the curvature mass, we evaluate the zero momentum self-energy, $\Pi_{ab}^{(X)}\equiv\Pi_{ab}^{(X)}(Q=0),$ expressing it in terms of the integrals 
\be
\begin{split}
I^X(m_1,m_2)&\equiv I^X(Q=0;m_1,m_2),\\
I^X(m)&\equiv I^X(Q=0;m).
\end{split}
\label{Eq:I_X_def_zero_mom}
\ee
These are calculated in Appendix~\ref{app:PiCalcDetails}, where, using partial fractioning and simple algebraic manipulations, they are reduced to a combination of simple integrals.

In the case of \hbox{(pseudo)}scalars the result is summarized in Table~\ref{Tab:Pi_all}, where we indicate the quark masses, labeled by $f$ and $f'$, to be used in the formula of the one-loop self-energy for a given choice of the flavor indices $a$ and $b$. The correction to the tree-level curvature mass $\hat m^{2,(S/P)}_{ab}$ is of the form
\be
\begin{aligned}
  &\hat M^{2,(S/P)}_{ab} = \hat m^{2,(S/P)}_{ab} + \Delta\hat m^{2,(S/P)}_{ab} + \delta\hat m^{2,(S/P)}_{ab}, \\
  &\Delta\hat m^{2,(S/P)}_{ab} \equiv \Pi_{ab,\textrm{vac}}^{(S/P)},\qquad  \delta\hat m^{2,(S/P)}_{ab} \equiv \Pi_{ab,\textrm{mat}}^{(S/P)},
\end{aligned}
\label{Eq:M2SP_gen_form}
\ee
where the \emph{vacuum} part needs renormalization and the \emph{matter} part is finite and determined by ${\cal T}^{(1)}_f$ (and its mass derivative, for some flavor indices). In some flavor cases Eq.~\eqref{Eq:M2SP_gen_form} does not represent the physical curvature mass of the \hbox{(pseudo)}scalars, due to their mixing with \hbox{(axial-)}vectors. This issue is addressed in Sec.~\ref{sec:mixing}, where we will see that the mixing affects all the pseudoscalars, but only the scalars with flavor indices $4-7.$

In the case of the \hbox{(axial-)}vectors, the evaluation of the self-energy requires some care. The self-energy is split into \emph{vacuum} and \emph{matter} parts, as indicated in Table~\ref{Tab:Pi_all}. For some flavor indices, namely $a=3,{\rm N,S}$ for $\phi_3\ne0$ and additionally $a=1,2$ for $\phi_3=0$, the \emph{vacuum} part of the vector self-energy requires as in the $N_f=1$ case, a covariant calculation in a regularization scheme that complies with the requirement $\Pi^{(V),\mu\nu}_{\rm vac}(Q=0)=0,$ which is familiar from QED. This requirement is investigated  in Appendix~\ref{app:PiV_prop}, where we relate it to a symmetry of the classical Lagrangian, which is manifest for a specific field background. 

For simplicity, we use dimensional regularization to calculate the \hbox{(axial-)}vector self-energy, irrespective of the flavor index. The \emph{vacuum} integral determining the self-energy can be reduced to tadpole integrals (see Eq.~\eqref{Eq:I_VA_tad_reduced}). Its finite and divergent pieces are given in Eqs.~\eqref{Eq:V_A_2masses_vac_result} and \eqref{Eq:V_A_eq_masses_vac_result}. For the \emph{matter} part we only need to consider purely temporal ($00$) and spatial ($ij$) components, as mixed ($0i$) components vanish due to symmetric integration. The \emph{matter} part of the relevant integrals, given in Eqs.~\eqref{Eq:IAV_00_mat} and \eqref{Eq:IAV_11_mat}, contains also an integral whose mass derivative is proportional to the tadpole, $\frac{d{\cal U}^{(1)}_f}{d m_f^2}=-\frac{3}{2} {\cal T}_f^{(1)}$ (see Eq.~\eqref{Eq:rel_dUdm2-T}). In the equal mass limit this relation considerably simplifies the result.  

A further complication with the \hbox{(axial-)}vectors is related to the fact that one needs to consider the mode decomposition of the dressed propagator. This is done in Appendix~\ref{app:decomp}, using the usual set of tensors that includes three- and four-dimensional projectors. As shown there, each mode has its own one-loop curvature mass, determined by the Lorentz components of the self-energy tensor in the $Q\to0$ limit.

Using the form of the self-energy given in Table~\ref{Tab:Pi_all} in the expression \eqref{Eq:PiVA_comp} that gives the contribution to a given mode $\rm p\in\{t,l,L\}$, one sees that the curvature mass has the structure
\be
\hat M^{2,(V/A)}_{i,\rm p}=\hat m_i^{2,(V/A)} + \Delta \hat m_i^{2,(V/A)} + \delta_{\rm p}\hat  m_i^{2,(V/A)},
\label{Eq:M2VA_gen_form}
\ee
where $i$ refers either to flavor indices ({\it e.g.} $ab=44$) or to the particle ({\it e.g.} $K_1$). $\Delta \hat m_i^{2}$ is the contribution of the \emph{vacuum} part $\propto I^{V/A}_{\rm vac}$, which in view of Eq.~\eqref{Eq:PiVA_comp} is the same for all modes. $\delta_{\rm p}\hat  m_i^{2}$ is the mode-dependent contribution of the \emph{matter} part, and it is determined by $\propto I^{V/A,00}_{\rm mat}$ for the `$\rm l$' mode and $\propto I^{V/A,11}_{\rm mat}$ for the `$\rm t$' and `$\rm L$' modes, as discussed around Eqs.~\eqref{Eq:nzT_PiV_comp_Lor-ind_rel_eq_mass} and \eqref{Eq:nzT_PiA_comp_Lor-ind_rel}.

The `$\rm t$' and `$\rm l$' modes are, respectively, 3-transverse and 3-longitudinal, while the `$\rm L$' mode is 4-longitudinal. We will see in Sec.~\ref{sec:mixing} that the `${\rm L}$' mode \eqref{Eq:M2VA_gen_form} influences the physical curvature mass of the (pseudo)scalars.

For the flavor indices appearing in the last three rows of Table~\ref{Tab:Pi_all} (and also for $ab=11,22$ for $\phi_3=0$, when $m_u=m_d$) one has only a matter fermionic contribution to the vector curvature mass and only in the case of the '$\rm l$' mode. This is because the single mass integral is such that $I^V_{\rm vac}(m_f)=I^{V,11}_{\rm mat}(m_f)=0,$ as shown in Appendix~\ref{app:PiCalcDetails}. In the isospin symmetric case the matter contributions to the curvature mass of the modes are (note that due to the absence of mixing $\omega_{\rm N}\equiv \omega(782)$ and  $\omega_{\rm S}\equiv \phi(1020)$)
\bea
  &&\delta_t\hat m^2_i=\delta_L\hat m^2_i = 0, \qquad i=\rho,\omega_{\rm N}, \omega_{\rm S},\nonumber\\
  &&\delta_{\rm l}\hat m^2_{\rho/\omega_{\rm N}}=C_V I^{V,00}_{\rm mat}(m_l), \quad \delta_{\rm l}\hat m^2_{\omega_{\rm S}}=C_V I^{V,00}_{\rm mat}(m_s),\nonumber\\
  &&\delta_{\rm t,L}\hat m^2_{K^\star}=-\Pi^{(V),11}_{44,\rm mat}=-C_V I^{V,11}_{\rm mat}(m_l,m_s),\nonumber\\
  &&\delta_{\rm l}\hat m^2_{K^\star}=\Pi^{(V),00}_{44,\rm mat}=C_V I^{V,00}_{\rm mat}(m_l,m_s)
\eea
for the vectors and
\bea
&&\delta_{\rm t,L}\hat m^2_{a_1/f_{1\rm N}}=-\Pi^{(A),11}_{11,\rm mat}=-C_V I^{A,11}_{\rm mat}(m_l),\nonumber\\
&&\delta_{\rm l}\hat m^2_{a_1/f_{1\rm N}}=\Pi^{(A),00}_{11,\rm mat}=C_V I^{A,00}_{\rm mat}(m_l),\nonumber\\
&&\delta_{\rm t,L}\hat m^2_{K_1}=-\Pi^{(A),11}_{44,\rm mat}=-C_V I^{A,11}_{\rm mat}(m_l,m_s),\nonumber\\
&&\delta_{\rm l}\hat m^2_{K_1}=\Pi^{(A),00}_{44,\rm mat}=C_V I^{A,00}_{\rm mat}(m_l,m_s),
\eea
for the axial-vectors, where $C_V=2N_cg_V^2$ and for the $f_{1\rm S}$ meson the contributions are as for $f_{1\rm N}$ with $m_l$ replaced by $m_s.$ The integrals are explicitly given in Appendix~\ref{app:PiCalcDetails}.

The vacuum contributions need renormalization and their finite part is given for $\phi_3=0$ in Eqs.~\eqref{Eq:vac_V_curv_m2} and \eqref{Eq:vac_A_curv_m2}.

\subsection{Connection to previous calculations \label{ss:prev-calc}}

The fermionic correction to the (pseudo)scalar curvature masses was calculated first in Ref.~\cite{Schaefer:2008hk} in the isospin symmetric case ($\phi_3=0$). The Polyakov-loop degrees of freedom were incorporated in Ref.~\cite{Gupta:2009fg}. Bringing the expression in Eq.~(B12) of \cite{Schaefer:2008hk} and in Eq.~(25) of \cite{Gupta:2009fg} in a form containing the tadpole and the bubble integral at vanishing momentum is not mandatory, but it reveals the structure behind the obtained result for the curvature mass. Also, integration by parts shows that the result can be given in terms of a single function: the Fermi-Dirac distribution or the modified Fermi-Dirac distribution \eqref{Eq:mod_FD}, when Polyakov-loop degrees of freedom $\Phi$ and $\bar\Phi$ are included. This simple observation makes superfluous the introduction of $B^\pm_f$ and $C^\pm_f,$ used also in \cite{Kovacs:2016juc} following \cite{Gupta:2009fg}, and allows for a slight simplification of the formulas used so far in the literature.

Using the method of Ref.~\cite{Schaefer:2008hk} we show below how to obtain the expressions of the (pseudo)scalar curvature masses given in Table~\ref{Tab:Pi_all}. The method assumes that in the ideal gas contribution of the three quarks to the grand potential we can use quark masses that depend on the fluctuating mesonic fields, as in Eq.~\eqref{Eq:U_f} of the $N_f=1$ case. The method works because for $g_V=0$ and $K=0$ the eigenvalues of the mass matrix in Eq.~\eqref{Eq:matrix_Nf3} correspond to the $u, d,$ and $s$ quark sectors. In case of the \hbox{(axial-)}vectors, it is not enough to concentrate on the mass matrix, as explained in Sec.~\ref{sec:Intro}. Taking \hbox{(axial-)}vector field derivatives of the eigenvalues of the mass matrix, as in Ref.~\cite{Tawfik:2014gga}, results in a curvature mass tensor which breaks Lorentz covariance, as it is not proportional to $g_{\mu\nu}$ at $T=0$.

Concentrating on the matter part of the grand potential, we start from its expression given in the ideal gas approximation in Eq.~(27) of \cite{Kovacs:2016juc}
\begin{equation}
  \Omega_{\bar q q}^{(0)\textnormal{T}}(T,\mu_q) = -2 T \sum_{f=u,d,s}\int\frac{d^3 p}{(2\pi)^3}\big[\ln g_f^+(p) + \ln g_f^-(p)\big],\ \ 
\label{Eq:fermi_omega}
\end{equation}
where
\begin{equation}
  g_f^{\pm}(p) = 1 + 3\left( \Phi^\pm + \Phi^\mp e^{-\beta E_f^\pm(p)}
  \right) e^{-\beta E_f^\pm(p)} + e^{-3\beta  E_f^\pm(p)},
\label{Eq:g_Phi_barPhi}
\end{equation}
with $\Phi^+ = \bar\Phi, \Phi^-=\Phi$, $E_f^{\pm}(p) = E_f(p)\mp\mu_f$ and $E_f^2(p)=p^2+M_f^2.$ Here $M_f$ are the eigenvalues of the matrix in Eq.~\eqref{Eq:matrix_Nf3}, which depend not only on the scalar background, but also on the fluctuating (pseudo)scalar fields, generically denoted by $\varphi_a$, with $a$ being the flavor index. After taking the second derivative with respect to $\varphi_a$, the fluctuating field is set to zero, in which case $M_f(\varphi_a=0)=m_f$.

The modified Fermi-Dirac distribution functions
\begin{equation}
  F^\pm_f(p) =\frac{ \Phi^\pm e^{-\beta E_f^\pm(p)} + 2\Phi^\mp e^{-2 \beta E_f^\pm(p)} + e^{-3\beta E_f^\pm(p)} } {g^\pm_f(p)}
  \label{Eq:mod_FD}
\end{equation}
are introduced by an integration by parts
\bea
T \int\frac{d^3 p}{(2\pi)^3} \ln g_f^\pm(p) 
= \frac{1}{2\pi^2}\int_0^\infty d p p^4 \frac{F^\pm_f(p)}{E_f(p)}.
\eea

Then one uses that the dependence on $\varphi_a$ is through $M_f^2,$ which only appears in the combination $p^2+M_f^2,$
\be
\frac{\partial}{\partial \varphi_a}\frac{F^\pm_f(p)}{E_f(p)} =
\frac{1}{2 p} \frac{\partial}{\partial p}\bigg(\frac{F^\pm_f(p)}{E_f(p)}\bigg)\frac{\partial M_f^2}{\partial \varphi_a}.
\ee
The above relation and integration by parts results in
\bea
\frac{\partial^2 \Omega_{\bar q q}^{(0)\textnormal{T}}(T,\mu_q)}{\partial\varphi_a \partial \varphi_b}\bigg|_{\varphi=0}
&=& -6 \sum_f\bigg[\frac{\partial^2 M_f^2}{\partial\varphi_a \partial \varphi_b}{\cal T}_f^{(1)}\nonumber\\
  && -  \frac{\partial M_f^2}{\partial \varphi_a} \frac{\partial M_f^2}{\partial \varphi_b} {\cal B}_f^{(1)}\bigg]_{\varphi=0},
\label{Eq:2nd_field_der_of_grand_pot}
\eea
where the integral ${\cal T}_f^{(1)}\equiv{\cal T}^{(1)}(m_f)$ and its mass derivative, defined in Eqs.~\eqref{Eq:tad_FD_T} and \eqref{Eq:Bub_T}, now contain the modified Fermi-Dirac distribution functions \eqref{Eq:mod_FD}.

Using Table~III of \cite{Schaefer:2008hk} for the derivatives of the masses (Table~II of \cite{Kovacs:2016juc} to also get the wave-function renormalization factors due to the shift of the (axial-)vector fields) one recovers the result obtained in \cite{Kovacs:2016juc} in the isospin symmetric case, where one has $m_{u,d}=m_l=g_S \phi_{\rm N}/2$. For example, in the $ab=11$ scalar sector, which is not affected by the mixing, one has ($M_s$ does not contribute)  
\be
\sum_{f=u,d}\frac{\partial^2 M_f^2}{\partial S_1 \partial S_1}\bigg|_{\varphi=0}
= g^2_S, \ \ 
\sum_{f=u,d}\frac{\partial M_f^2}{\partial S_1} \frac{\partial M_f^2}{\partial S_1}\bigg|_{\varphi=0}=2 g_S^2 m_l^2,
\ee
so that the \emph{matter} contribution of the fermions to the curvature mass obtained from Eq.~\eqref{Eq:2nd_field_der_of_grand_pot} has the form
\be
\delta m^2_{a_0}=-6 g_S^2 \big[{\cal T}_l^{(1)} - 2 m_l^2 {\cal B}_l^{(1)}\big],
\ee
in accordance with Table~\ref{Tab:Pi_all} in view of \eqref{Eq:Bub_T}.

The above simple calculation shows that in the presence of Polyakov degrees of freedom the fermionic contribution to the curvature mass can be given in terms of the modified Fermi-Dirac distribution functions \eqref{Eq:mod_FD}. Based on this, one can safely replace in our previous matter integrals $f^\pm_f(p)$ with $F^\pm_f(p).$
 
\section{Renormalization of the curvature mass \label{sec:ren}}

For simplicity, we discuss the renormalization of the fermionic correction to the curvature masses only in the isospin symmetric case ($\phi_3=0$) where $m_l\equiv m_u=m_d.$ Then, according to Table~\ref{Tab:flavor_tr}, the contribution in the last row of Table~\ref{Tab:Pi_all} vanishes, while for $1-3,N$ flavor indices one has to use the equal mass formula of the $a={\rm S}$ case with the replacement $m_s\to m_l$.

Since the renormalization of the \hbox{(pseudo)}scalar curvature masses poses no problem and was already done in the literature, using dimensional regularization \cite{Schaefer:2011ex, Chatterjee:2011jd, Tiwari:2013pg} or cutoff regularization \cite{Kovacs:2016juc}, we will only sketch an alternative method, which can be used in a localized approximation, that is when the self-energy is evaluated at vanishing momentum.

The divergence(s) of a \emph{vacuum} integral can be separated by expanding a localized propagator around the auxiliary function $G_0(K)=1/(K^2-M_0^2),$ where $M_0$ plays the role of a renormalization scale. For the tadpole integral, iterating once the identity $G_f = G_0 + (m_f^2-M_0^2) G_0 G_f,$ one obtains upon integration
\be
{\cal T}^{(0)}(m_f) = D_2 + m_f^2 D_0 + {\cal T}_\textrm{F}^{(0)}(m_f),
\label{Eq:Tad_felbont}   
\ee
where the first and second terms are the overall divergence and the subdivergence given in terms of
\be
D_2={\cal T}^{(0)}(M_0) - M_0^2 D_0,  \quad D_0=\frac{d {\cal T}^{(0)}(M_0)}{d M_0^2},
\ee
and the last term in Eq.~\eqref{Eq:Tad_felbont} is finite, 
\bea
{\cal T}_\textrm{F}^{(0)}(m_f)&=&i(m_f^2-M_0^2)^2\int_K G_0^2(K)G_f(K)\nonumber \\
&=& \frac{1}{16\pi^2}\bigg(M_0^2-m_f^2+m_f^2\ln\frac{m_f^2}{M_0^2}\bigg).
\label{Eq:Tad_vac_F}
\eea

With the above renormalization procedure the finite part of the tadpole is independent of whether covariant or noncovariant calculation, cutoff or dimensional regularization is used (provided the cutoff is sent to infinity in Eq.~\eqref{Eq:Tad_vac_F}). In a noncovariant calculation, Eq.~\eqref{Eq:Tad_vac_F} is obtained from Eq.~\eqref{Eq:tad_vac_noncov} by writing $E_f(k)=(k^2+M_0^2+\Delta m_f^2)^{1/2}$, with $\Delta m_f^2=m_f^2-M_0^2,$ and subtracting from the \emph{vacuum} piece of the tadpole the first two terms obtained by expanding $1/E_f(k)$ in powers of $\Delta m^2_f$. Subtracting also the ${\cal O}\big((\Delta m^2_f)^2\big)$ term when renormalizing the integral
\be
   {\cal L}^{(0)}(m_f)=\int\frac{d^3 k}{(2\pi)^3} E_f(k),
\label{Eq:tr-log_int_nonconv}   
\ee
which determines the one-loop fermionic contribution to the effective potential in Eq.~\eqref{Eq:Veff_MF} (and, with the replacement $m_f^2\to \hat M^2$, also the contribution of the ring integrals with localized self-energy in Eq.~\eqref{Eq:Veff_GA}), results in the following finite \emph{vacuum} part
\be
   {\cal L}^{(0)}_{\rm F}(m_f)=-\frac{1}{64\pi^2}\bigg[
     \Delta m_f^2\big(2m_f^2+\Delta m_f^2\big)-2m_f^4 \ln\frac{m_f^2}{M_0^2}    
     \bigg],
\label{Eq:tr-log_int_nonconv_ren}   
\ee
which satisfies $d {\cal L}^{(0)}_{\rm F}(m_f)/d m_f^2 = {\cal T}^{(0)}_{\rm F}(m_f),$ that is, the relation also holding between the unrenormalized integrals \eqref{Eq:tr-log_int_nonconv} and \eqref{Eq:tad_vac_noncov}.

\medskip
Now we turn our attention to the renormalization of the \hbox{(axial-)}vector curvature masses \eqref{Eq:M2VA_gen_form}. The relevant terms of the EL$\sigma$M Lagrangian introduced in Eq.~(2) of Ref.~\cite{Parganlija:2012fy} are those proportional to the coupling $h_i,i=1,2,3$ and the term containing the covariant derivative. In dimensional regularization, used here with $d=4-2\epsilon,$ no overall divergence is encountered, and thus the mass term of the \hbox{(axial-)}vectors $\propto m_1^2$ is not needed. The tree-level mass squared of the vector and axial-vector fields depend on the strange and nonstrange scalar condensates $\phi_N$ and $\phi_S$, as a result of the coupling of these fields to the scalars, which acquire an expectation value. We have to ensure that the subdivergence of the \emph{vacuum} contribution to the curvature mass in Eq.~\eqref{Eq:M2VA_gen_form} is removed by the environment-dependent terms (that is, proportional with $\phi_N$ and $\phi_S$) present in the tree-level mass formulas.

Using Eqs.~\eqref{Eq:V_A_2masses_vac_result} and \eqref{Eq:V_A_eq_masses_vac_result} in the expressions of Table~\ref{Tab:Pi_all} with $m_{u,d}=m_l$, we see that the \emph{vacuum} piece of the vector curvature mass is divergent only for flavor indices $4-7$, corresponding to the $K^\star$ meson,
\be
\Delta\hat m^2_{K^\star,\textrm{div}} = g_V^2 N_c \frac{(m_s - m_l)^2}{16\pi^2\epsilon},
\ee
while for the axial-vectors divergence is present in all the flavor sectors
\bea
\Delta\hat m^2_{a_1,\textrm{div}} &=& \Delta\hat m^{2,(A)}_\textrm{NN, div}= g_V^2 N_c \frac{4 m_l^2}{16\pi^2\epsilon},\nonumber \\
\Delta\hat m^{2,(A)}_\textrm{SS, div} &=& g_V^2 N_c \frac{4 m_s^2}{16\pi^2\epsilon},\nonumber \\
\Delta\hat m^2_{K_1.\textrm{div}} &=& g_V^2 N_c \frac{(m_s + m_l)^2}{16\pi^2\epsilon},
\eea
where the $a_1$ ($K_1$) meson correspond to the $1-3$ ($4-7$) flavor indices.

The above subdivergence structure means that in the tree-level mass formulas given in \cite{Parganlija:2012fy} in Eqs.~(27)-(34), we have to look for terms which are only present for $K^\star$ and the axial-vectors. There is indeed such a term, the one proportional with $g_1^2$, and by using the mass formulas \eqref{Eq:m_f_tree} of the quarks and also $(m_l\pm m_s)^2=g_S^2\left(\frac{\phi^2_{\N}}{4}\pm\frac{\phi_{\N}\phi_{\Ss}}{\sqrt{2}}+\frac{\phi_{\Ss}^2}{2}\right)$, we see that the environment-dependent part in the tree-level mass formulas matches the form of the subdivergence, which therefore can be removed. The only problem is that, since $g_1$ is squared, absorbing the subdivergence in the counterterm of $g_1$ would result in an awkward renormalization, as the term quadratic in the counterterm should be dropped.

A close inspection of the structure of the terms in Eqs.~(27)-(34) of \cite{Parganlija:2012fy} shows that one can achieve renormalization by assigning counterterms to the couplings $h_i,i=1,2,3$ instead of $g_1$. Namely, splitting the bare coupling into renormalized one and counterterms, $h_i=h_{iR}+\delta h_i,$ one sees that the subdivergences can be eliminated with the counterterms:
\[
\delta h_1=0\quad \textrm{and}\quad \delta h_2=-\delta h_3=-\frac{N_cg_S^2 g_V^2}{16\pi^2\epsilon}.
\]

The fact that renormalization can be achieved without referring to the counterterm of $g_1$ raises the question of why $g_1^2$ is present at all in the tree-level mass formulas. A closer look into the origin of the terms proportional to $h_2, h_3,$ and $g_1^2$ in the mass formulas reveals that some terms included in the Lagrangian through the terms proportional to $h_2$ and $h_3$ are also generated by the covariant derivative term which contains $g_1^2$. Namely, using the covariant derivative  of \cite{Parganlija:2012fy},
\be
D^\mu M = \partial^\mu M - i g_1 ( L^\mu M - M R^\mu) - i e A_e^\mu[T_3,M],
\ee
where $M=S+i P$ as in \cite{Kovacs:2016juc}, the coefficient of the ${\cal O}(g_1^2$) term in $\Tr\big[(D_\mu M)^\dagger\big(D_\mu M)]$ is
\be
 \Tr( L\cdot L M M^\dagger + R\cdot R M^\dagger M) - \Tr[2 L_\mu M R^\mu M^\dagger].
\ee
The above two traces were added to the Lagrangian with coefficients $h_2$ and $h_3$, respectively. Therefore, using $L_\mu^\dagger=L_\mu$ and the shorthand $|L_\mu M|^2\equiv\big(L_\mu M\big)^\dagger \big(L^\mu M\big)$, the Lagrangian used in \cite{Parganlija:2012fy} is in fact 
\be
\delta\mathcal{L}_{\tilde h_{2,3}}= \tilde h_2\Tr(|L_\mu M|^2 + |R_\mu M|^2) + 2 \tilde h_3 \Tr[L_\mu M R^\mu M^\dagger],
\ee
where the relations between the parameters are
\be
h_2 = \tilde h_2 - g_1^2\quad \textrm{and}\quad  h_3 = \tilde h_3 + g_1^2,
\ee
from which $h_2 + h_3 = \tilde h_2 + \tilde h_3$. Applying these relations, $g_1^2$ can be eliminated from the tree-level mass formulas of the \hbox{(axial-)}vectors in which $h_{2,3}$ is replaced by $\tilde h_{2,3}$.

To avoid duplication of terms in the Lagrangian, it is a better practice to use a covariant derivative containing only the electromagnetic field,
\be
\bar D^\mu M = \partial^\mu M - i e A_e^\mu[T_3,M],
\ee
and write the Lagrangian that contains the mass terms of the \hbox{(axial-)}vectors and their interaction with the \hbox{(pseudo)}scalars in the form 
\bea
\delta{\cal L} &=& \Tr\big[(\bar D_\mu M)^\dagger\big(\bar D_\mu M)] \nonumber\\
&+& \frac{1}{2}\big[m_1^2 + h_1\tr\big(M^\dagger M\big)\big]\tr\big(L\cdot L + R\cdot R \big) + \delta\mathcal{L}_{\tilde h_{2,3}}\nonumber\\
&+& g_1 \Tr\big[i(M^\dagger L_\mu - R_\mu M^\dagger)(\bar D^\mu M) + \textrm{h.c.} \big],
\label{Eq:proposed_L}
\eea
although this form is less compact than the one in \cite{Parganlija:2012fy}.

After all these considerations, we give for completeness the vacuum curvature masses containing the renormalized one-loop level contribution of the fermions. The vector curvature masses are
\be
\begin{aligned}
  & \hat M^2_{\rho,{\rm vac}} = \hat M^2_{\omega {\rm N},{\rm vac}} = \hat m^2_\rho = \hat m^2_{\omega {\rm N}}, \\
  &\hat M^2_{\omega {\rm S}, {\rm vac}} = \hat m^2_{\omega {\rm S}},  \\
  & \hat M^2_{K^\star,{\rm vac}} = \hat m^2_{K^\star} + 2 N_c g_V^2 I^V_{\rm vac, F}(m_l,m_s),
\end{aligned}
\label{Eq:vac_V_curv_m2}
\ee
while the axial-vectors ones are
\be
\begin{aligned}
  & \hat M^2_{a_1/f_{\rm 1N},{\rm vac}}= \hat m^2_{a_1/f_{\rm 1N}} + 2 N_c g_V^2 I^A_{\rm vac, F}(m_l),\\
  & \hat M^2_{f_{\rm 1S},{\rm vac}}= \hat m^2_{f_{\rm 1S}} + 2 N_c g_V^2 I^A_{\rm vac, F}(m_s),\\
  & \hat M^2_{K_1,{\rm vac}}= \hat m^2_{K_1} + 2 N_c g_V^2 I^A_{\rm vac, F}(m_l,m_s),
\end{aligned}
\label{Eq:vac_A_curv_m2}
\ee
where the classical curvature masses
\be
\begin{aligned}
  &\hat m^2_{\rho/a_1}=m_1^2+\frac{H_\pm}{2}\phi^2_{\rm N}+\frac{h_{\rm 1R}}{2}\phi^2_{\rm S},\\
  &\hat m^2_{K^\star/K_1}=m_1^2+\frac{H_1}{4}\phi_{\rm N}^2\pm\frac{h_{\rm 3R}}{\sqrt{2}}\phi_{\rm N}\phi_{\rm S} + \frac{H_2}{2}\phi_{\rm S}^2,\\
  &\hat m^2_{\omega_{\rm S}/f_{\rm 1S}}=m_1^2+\frac{h_{\rm 1R}}{2}\phi^2_{\rm N}+\bar H_\pm\phi^2_{\rm S}
\end{aligned}
\ee
are written, omitting the symmetry breaking terms considered in \cite{Parganlija:2012fy}, using the constants $H_\pm=h_{\rm 1R}+h_{\rm 2R}\pm h_{\rm R3},$ $H_1=2h_{\rm 1R}+h_{\rm 2R},$ $H_2=h_{\rm 1R}+h_{\rm 2R},$ and $\bar H_{\pm}=\pm h_{\rm 3R} + h_{\rm 2R} + h_{\rm 1R}/2.$

\section{$S-V$ and $P-A$ mixing \label{sec:mixing}}

These types of mixings come from the last line of Eq.~\eqref{Eq:proposed_L} after performing the trace and shifting the scalar fields with their background values. Doing also a symmetrization using integration by parts in the classical action, one obtains, in Fourier space and at quadratic order in the fluctuating fields, the last four mixing (crossed) terms in Eq.~(9) of \cite{Parganlija:2012fy} (up to an unnecessary factor of $i$ in the $V-S$ mixing terms and the wrong sign of the last two terms):
\bea
\delta\mathcal{L}_{g_1}^\textrm{quad}&=&-\frac{g_1}{2} i K_\mu\Big[d_{ijk}\big(\tilde A_i^\mu\bar P_j-\tilde P_i\bar A_j\big)\nonumber\\
  &+& f_{ijk}\big(\tilde V_i^\mu \bar S_j+ \tilde S_i\bar V_j^\mu\big)\Big]\phi_k,\ \  i,j,k=0,\dots,8,\ \quad
\label{Eq:L_mixing}
\eea
where $\tilde X\equiv \tilde X(K)$ and $\bar X \equiv \tilde X(-K)$.

Due to the values of $f_{ijk}$, the $S-V$ mixing occurs only in the $4-5$ and $6-7$ flavor sectors and is of the form $S_{a/b}-V_{b/a}$ with $ab=45$ and $ab=67,$ respectively. The $P-A$ mixing occurs in all flavor sectors and has the form $P_a-A_a$ for $a\ne 3,N,S$. The $P-A$ mixing in the $3-{\rm N}-{\rm S}$ flavor sectors simplifies in the isospin symmetric case ($\phi_3=0$), but, nevertheless, it involves an additional ${\rm N}-{\rm S}$ mixing in the pseudoscalar sector.

At the classical level, the usual way to eliminate the mixing term is by shifting (in direct space) the \hbox{(axial-)}vector field by the derivative of the \hbox{(pseudo)}scalar field with an appropriately chosen wave-function renormalization constant as prefactor \cite{Gasiorowicz:1969kn, Pisarski:1994yp, Parganlija:2012fy}.

Here we adopt a different strategy and show that the wave-function renormalization constant is recovered when one identifies the contribution of the physical modes to the partition function evaluated in the ideal gas approximation, discussed in Sec.~\ref{sec:GA}. In this approximation the bosonic fluctuations are neglected in the fermionic determinant obtained by integrating out the fermions in the partition function and, keeping only quadratic terms in the mesonic Lagrangian, the Gaussian functional integral is done over the \hbox{(axial-)}vector and \hbox{(pseudo)}scalar fields. Then, we apply the same method at finite $T$ in the Gaussian approximation, that is when the quadratic part of the mesonic Lagrangians is corrected by the field expansion of the fermionic determinant. Considering self-energies at vanishing momentum, we find that the form of the wave-function renormalization constant, resulting from the mixing of the \hbox{(pseudo)}scalar with the nonpropagating 4-longitudinal \hbox{(axial-)}vector mode, is unchanged at $T\ne0$, only that it involves one-loop curvature masses instead of the tree-level ones that appear in the ideal gas approximation.

\subsection{Classical level mixing 
\label{ss:cl-mix}
}

\subsubsection{$S-V$ mixing}
We start with the mixing in the $4-5$ flavor sectors. Using Eq.~\eqref{Eq:L_mixing} and Eq.~(9) of Ref.~\cite{Parganlija:2012fy} one obtains
\be
\delta \mathcal{L}^{SV}_{45}=\frac{1}{2}\bigg[(\tilde S_4,\tilde V_5^\mu) \mathds{M}^{45}_{\mu\nu}\m{\bar S_4\\\bar V_5^\nu}+(\tilde S_5,\tilde V_4^\mu) \mathds{M}^{45*}_{\mu\nu} \m{\bar S_5\\\bar V_4^\nu} \bigg],
\label{Eq:LSV45}
\ee
where the $5\times 5$ matrix is
\be
\mathds{M}^{45}_{\mu\nu} = \m{i \mathcal{D}^{-1}_{4}(K) & -i c_{45} K_\nu\\ i c_{45} K_\mu & i\mathcal{D}^{-1}_{\mu\nu,5}(K)},
\label{Eq:mixM_VS45}
\ee
with $c_{45}= g_1f_{45k}\phi_k=g_1(\phi_3+\phi_{\rm N}-\sqrt{2}\phi_{\rm S})/2$.
The tree-level inverse propagators are ($\hat m^2_{K^{\star\pm}}\equiv \hat m^{2,(V)}_{44}$)
\bea
\label{Eq:inv_prop_in_mixing_matrix}
&&i\mathcal{D}^{-1}_{4}(K)\equiv i\mathcal{D}^{-1}_{5}(K)=K^2-\hat m^{2,(S)}_{44},\\
&&i\mathcal{D}^{-1}_{\mu\nu,4}(K)\equiv i\mathcal{D}^{-1}_{\mu\nu,5}(K)=\hat m^2_{K^{\star\pm}} {\rm P}^{\rm L}_{\mu\nu} + \big(\hat m^2_{K^{\star\pm}}-K^2\big){\rm P}^{\rm T}_{\mu\nu},\nonumber
\eea
where we used the usual 4-longitudinal and 4-transverse projectors ${\rm P}^{\rm L/T}_{\mu\nu}$ (see Eq.~\eqref{Eq:PLT_zeroT}) and that $\hat m^{2,(S/V)}_{44}=\hat m^{2,(S/V)}_{55}.$

Doing in the partition function the Gaussian integral over the fields appearing in Eq.~\eqref{Eq:LSV45} one obtains
\be
\ln Z^{(2)}_{VS,45} = -\frac{1}{2}\int_K \ln\det  \mathds{M}^{45}_{\mu\nu} - \frac{1}{2}\int_K \ln\det  \mathds{M}^{45*}_{\mu\nu}.
\label{Eq:Z_from_LSV45}
\ee
The calculation is simplified by the identity
\be
\det\m{A & B\\C &D} = \det(A) \det\big(D - C A^{-1} B\big),
\label{Eq:det_identity}
\ee
which gives in the present case 
\bea
\det \mathds{M}^{45}_{\mu\nu} &=& \det \mathds{M}^{45*}_{\mu\nu}= i \mathcal{D}_{4}^{-1}(K)\nonumber\\
&\times & \det\big(i \mathcal{D}^{-1}_{\mu\nu,4}(K) + i c^2_{45} \mathcal{D}_{4}(K) K^2\, {\rm P}^{\rm L}_{\mu\nu} \big).\ \
\eea
Using Eq.~\eqref{Eq:inv_prop_in_mixing_matrix} we have the projector decomposition of the remaining $4\times 4$ matrix, hence computing its determinant is an easy task. Given that ${\rm P}_{\rm T\mu}^\mu=3{\rm P}_{\rm L\mu}^\mu=3$, one obtains
\bea
\det \mathds{M}^{45}_{\mu\nu}=C^2_{45}\left(K^2-\hat m^2_{K_0^{\star\pm}}\right) \big(K^2-\hat m^2_{K^{\star\pm}}\big)^3,
\label{Eq:det_SV45_final}
\eea
where $C^2_{45}=\hat m^2_{K^{\star\pm}}-c^2_{45}$. The physical mass squared 
\bea
\hat m^2_{K_0^{\star\pm}}= Z^2_{K_0^{\star\pm}} \hat m^{2,(S)}_{44} \qquad  \textrm{with}\qquad Z^2_{K_0^{\star\pm}}=\frac{\hat m^2_{K^{\star\pm}}}{C^2_{45}},\quad
\label{Eq:m2scalar_Z2_SV45}
\eea
of the scalar mode arises as a result of its mixing with the nonpropagating 4-longitudinal vector mode. $Z_{K_0^{\star\pm}}$ is the wave-function renormalization constant.

The momentum-independent prefactor $C^2_{45}$ in Eq.~\eqref{Eq:det_SV45_final} reflects the existence of the nonpropagating 4-longitudinal vector mode. When dimensional regularization is used to perform the momentum integral in Eq.~\eqref{Eq:Z_from_LSV45}, the logarithm of the partition function receives contributions only from the propagating modes, represented by the two brackets in Eq.~\eqref{Eq:det_SV45_final}, {\it i.e.} there is no contribution from $\ln C^2_{45}$, which depends on the scalar backgrounds.

A similar calculation in the $6-7$ flavor sector gives
\bea
\det \mathds{M}^{67}_{\mu\nu}=C^2_{67}\left(K^2-\hat m^2_{K_0^{\star 0}}\right) \big(K^2-\hat m^2_{K^{\star 0}}\big)^3,
\eea
where $C^2_{67}=\hat m^2_{K^{\star 0}}-c^2_{67},$ with $\hat m^2_{K^{\star 0}}\equiv \hat m^{2,(V)}_{66}$ and $c_{67}= g_1f_{67k}\phi_k=g_1(-\phi_3+\phi_{\rm N}-\sqrt{2}\phi_{\rm S})/2.$ The physical scalar mass squared is
\be
\hat m^2_{K_0^{\star0}}= Z^2_{K_0^{\star 0}} \hat m^{2,(S)}_{66}, \qquad  Z^2_{K_0^{\star0}}=\hat m^2_{K^{\star0}}/C^2_{67}.
\ee

We mention that in the isospin symmetric case, $\phi_3=0$, one has $C^2_{67}=C^2_{45}$ and $\hat m^2_{K^{\star 0}}=\hat m^2_{K^{\star \pm}}$, hence $Z_{K_0^{\star\pm}} = Z_{K_0^{\star 0}}\equiv Z_{K_0^\star}$, as given in \cite{Parganlija:2012fy} in the last line of Eq.~(14).

\subsubsection{$P-A$ mixing}

We start with the $P_a-A_a, a=1,2$ mixing, given by
\begin{eqnarray*}
\delta \mathcal{L}^{PA}_{1\&2}=\frac{1}{2}\bigg[(\tilde P_1,\tilde A_1^\mu) \mathds{N}^{11}_{\mu\nu}\m{\bar P_1\\\bar A_1^\nu}+(\tilde P_2,\tilde A_2^\mu) \mathds{N}^{22}_{\mu\nu} \m{\bar P_2\\\bar A_2^\nu} \bigg],
\end{eqnarray*}
where the $5\times 5$ matrices are
\be
\mathds{N}^{11}_{\mu\nu} = \mathds{N}^{22}_{\mu\nu} = \m{i \mathcal{D}^{-1}_{1}(K) & i c_{11} K_\nu\\ -i c_{11} K_\mu & i\mathcal{D}^{-1}_{\mu\nu,1}(K)},
\label{Eq:N11_matrix}
\ee
with $c_{11}=g_1(\sqrt{2}\phi_0+\phi_8)/\sqrt{3}=g_1\phi_N$ and inverse propagators of the form given in Eq.~\eqref{Eq:inv_prop_in_mixing_matrix}, with masses $\hat m^{2,(P)}_{11}=\hat m^{2,(P)}_{22}$ and $\hat m^2_{a_1^{\pm}}\equiv \hat m^{2,(A)}_{11}=\hat m^{2,(A)}_{22},$ respectively.

The functional integral over $A_a,P_a, a=1,2$ and steps paralleling those leading to Eq.~\eqref{Eq:det_SV45_final} give ($C^2_{11}=\hat m^2_{a_1^{\pm}}-c^2_{11}$):
\be
\det \mathds{N}^{11}_{\mu\nu}=C^2_{11}\left(K^2-\hat m^2_{\pi^{\pm}}\right) \big(K^2-\hat m^2_{a_1^{\pm}}\big)^3,
\label{Eq:det_PA44_final}
\ee
with the physical mass of the pseudoscalar mode and the associated wave-function renormalization constant:
\be
\hat m^2_{\pi^{\pm}} = Z^2_{\pi^{\pm}} \hat m^{2,(P)}_{11} \quad \textrm{and} \quad
Z^2_{\pi^{\pm}}=\hat m^2_{a_1^{\pm}}/C^2_{11}.
\ee

A similar calculation involving fields with flavor indices $a=4,5$ and $a=6,7$ gives a determinant as in Eq.~\eqref{Eq:det_PA44_final}, with some obvious replacements:
\bea
&&C^2_{44}=\hat m^{2}_{K_1^{\pm}}-c^2_{44},\  \hat m^2_{K^{\pm}}= Z^2_{K^{\pm}} \hat m^{2,(P)}_{44},\  Z^2_{K^{\pm}}=\frac{\hat m^2_{K^{\pm}_1}}{C^2_{44}},\nonumber \\
&&C^2_{66}=\hat m^{2}_{K_1^{0}}-c^2_{66},\ \ \hat m^2_{K^{0}}= Z^2_{K^{0}} \hat m^{2,(P)}_{66},\ \ Z^2_{K^{0}}=\frac{\hat m^2_{K^{0}_1}}{C^2_{66}},\nonumber \\
\eea
where $c_{44/66}=g_1(\pm \phi_3+\phi_{\rm N}+\sqrt{2}\phi_{\rm S})/2.$ Again, for $\phi_3=0,$ one has a single wave-function renormalization constant, $Z_{K},$ given in Eq.~(13) of \cite{Parganlija:2012fy}.

Now we turn our attention to the mixing in the $3-{\rm N}-{\rm S}$ sectors, given by ($[P,A^\mu]_{ab}=\tilde P_a \bar A_b^{\mu} - \tilde A_b^\mu \bar P_a$)
\bea
\delta \mathcal{L}^{PA}_{3{\rm NS}}&=&\frac{i}{2}K_{\mu}\Big[c_{11}[P,A^\mu]_{33}+g_1\phi_3\big([P,A^\mu]_{3{\rm N}}+[P,A^\mu]_{{\rm N}3}\big)\nonumber \\
&&+c_{\rm NN}[P,A^\mu]_{{\rm NN}} + c_{\rm SS}[P,A^\mu]_{{\rm SS}} \Big],
\label{Eq:LPA3NS}
\eea
where $c_{\rm NN}=c_{11}=g_1\phi_{\rm N}$ and $c_{\rm SS}=g_1\sqrt{2}\phi_{\rm S}.$

For $\phi_3\ne0$ the complete quadratic Lagrangian involves a $15\times 15$ matrix. In this case, the appearance of the wave-function renormalization constant is nontrivial and will be presented elsewhere \cite{isobreak}. Here we consider the isospin symmetric case ($\phi_3=0$), in which the $(P_3,A_3^\mu)$ fields decouple. Their treatment is completely analogous to that of the $(P_1,A_1^\mu)$ sector, giving
\be
\det \mathds{N}^{33}_{\mu\nu}=C^2_{33}\left(K^2-\hat m^2_{\pi^0}\right) \big(K^2-\hat m^2_{a_1^0}\big)^3,
\ee
where $\hat m^2_{a_1^0}\equiv \hat m^{2,(A)}_{33}$ and
\be
C^2_{33}=\hat m^2_{a_1^0}-c^2_{11},\ \ \hat m^2_{\pi^0} = Z^2_{\pi^0} \hat m^{2,(P)}_{33}, \ \ Z^2_{\pi^0}=\frac{\hat m^2_{a_1^0}}{C^2_{33}},
\ee
with $\hat m^{2,(P/A)}_{33}=\hat m^{2,(P/A)}_{11}$ when $\phi_3=0$, and thus $Z_{\pi^0}=Z_{\pi^{\pm}}.$

The remaining $P-A$ mixing in the ${\rm N}-{\rm S}$ sectors is described by the $10\times 10$ matrix
\be
\mathds{N}^\textrm{NS}_{\mu\nu}=\m{\mathds{N}^\textrm{NN}_{\mu\nu} & \mathds{M} \\ \mathds{M} & \mathds{N}^\textrm{SS}_{\mu\nu}},\qquad \mathds{M}=\m{m^{2,(P)}_{\rm NS}& 0\\0 & 0},
\ee
where $\mathds{N}^\textrm{NN}_{\mu\nu}$ and $\mathds{N}^\textrm{SS}_{\mu\nu}$ are $5\times 5$ matrices of the form given in Eq.~\eqref{Eq:N11_matrix}, but with appropriate masses in the diagonal elements and constants $c_\textrm{NN/SS}$ in the off-diagonal ones.

The functional integral over the strange and nonstrange fields present in \eqref{Eq:LPA3NS} results in
\bea
&&\det \mathds{N}^{NS}_{\mu\nu}= C_\textrm{NN}^2 C_\textrm{SS}^2 (K^2-\hat m^2_{f_1{\rm N}})^3 (K^2-\hat m^2_{f_1{\rm S}})^3 \nonumber \\
&&\times \big[(K^2)^2 -(\hat m^2_{\eta{\rm N}} + \hat m^2_{\eta{\rm S}})K^2 + \hat m^2_{\eta{\rm N}}\hat m^2_{\eta{\rm S}} - (\hat m^2_{\eta{\rm NS}})^2 \big], \qquad
\label{Eq:det_NNS_matrix}
\eea
where $\hat m^2_{f_1a}=\hat m^{2,(A)}_{aa}$ and $C_{aa}^2=\hat m^2_{f_1a}-c^2_{aa}, a={\rm N},{\rm S}$. The classical pseudoscalar curvature masses used in Eq.~\eqref{Eq:det_NNS_matrix} contain the wave-function renormalization constants:
\be
\begin{aligned}
&\hat m^2_{\eta{\rm N}} = Z^2_{\rm N} \hat m^{2,(P)}_{\rm NN},& &Z^2_{\rm N} = \hat m^2_{f_1N}/C^2_{\rm NN},\\
&\hat m^2_{\eta{\rm S}} = Z^2_{\rm S} \hat m^{2,(P)}_{\rm SS},& &Z^2_{\rm S} = \hat m^2_{f_1S}/C^2_{\rm SS},\\
&\hat m^2_{\eta{\rm NS}} = Z_{\rm N} Z_{\rm S} \hat m^{2,(P)}_{\rm NS}.&&
\end{aligned}
\ee
In the second line of Eq.~\eqref{Eq:det_NNS_matrix}, one recognizes the elements of the mass squared matrix of the mixing $P_{\rm N}-P_{\rm S}$ sector. In terms of the physical eigenvalues of this matrix, namely
\be
\hat m^2_{{\eta'/\eta}}=\frac{1}{2}\big[\hat m^2_{\eta{\rm N}} + \hat m^2_{\eta{\rm S}}\pm\sqrt{(\hat m^2_{\eta{\rm N}} - \hat m^2_{\eta{\rm S}})^2-4(\hat m^2_{\eta{\rm NS}})^2} \big],
\ee
one obtains the final result,
\be
\det \mathds{N}^{NS}_{\mu\nu}=\prod_{a={\rm N,S}}C^2_{aa}(K^2-\hat m^2_{f_1a})^3
(K^2-\hat m^2_\eta)(K^2-\hat m^2_{\eta'}).
\ee
This contains the contribution of the propagating 4-transverse vector and physical pseudoscalar modes.

\subsection{Mixing in the Gaussian approximation \label{ss:GA-mix}}

We consider only the isospin symmetric case, $\phi_3=0$, in the {\it localized} approximation, in which the self-energies have vanishing momentum. In this case there is no correction to the off-diagonal elements of the $5\times 5$ matrices (also having Lorentz indices) considered in the previous subsection, while the diagonal elements are replaced by $i\mathcal{D}^{-1}(K)\to i\mathcal{G}_\textrm{loc}^{-1}(K)=K^2-\hat M^2,$ with $\hat M^2=\hat m^2+\Pi(0)$ and $i\mathcal{D}^{-1}_{\mu\nu}(K)\to i\mathcal{G}^{-1}_{\mu\nu,\rm{loc}}(K)=i\mathcal{D}^{-1}_{\mu\nu}(K)+\Pi_{\mu\nu}(0)$.

For the flavor indices involved in the mixing, vector and axial-vector self-energies have the same decomposition, given in Eq.~\eqref{Eq:nzT_tensor_decomp}. Basically what happens at $T\ne0$ is that in the inverse propagator the 4-transverse part encountered previously splits into 3-transverse and 3-longitudinal parts, with projectors $\textrm{P}^{\rm t}_{\mu\nu}$ and $\textrm{P}^{\rm l}_{\mu\nu}$, so that
\be
i\mathcal{G}^{-1}_{\mu\nu,\rm{loc}}(K)=\hat M^2_{\rm L} {\rm P}^{\rm L}_{\mu\nu}(K) +\sum_{{\rm p}={\rm l,t}}\big(\hat M^2_{\rm p}-K^2\big)\textrm{P}^{\rm p}_{\mu\nu}(K),
\label{Eq:inv_prop_loc_Gauss}
\ee
where $\hat M^2_{\rm l/t/L}=\hat m^2 + \Pi_{\rm l/t/L}(0)$. The components $\Pi_{\rm l/t/L}(0)$ are given in terms of the Lorentz components of the self-energy $\Pi_{\mu\nu}(0)$ in Appendix~\ref{app:decomp}, where details on the tensor decomposition can also be found.

Comparing Eq.~\eqref{Eq:inv_prop_loc_Gauss} to Eq.~\eqref{Eq:inv_prop_in_mixing_matrix} and using that $\det(L \textrm{P}_{\rm L} + l \textrm{P}_{\rm l} + t \textrm{P}_{\rm t}) = -L l t^2$ one immediately sees how to modify our previous results, obtained in the ideal gas approximation: instead of three 4-transverse \hbox{(axial-)}vector modes one has the contribution of two 3-transverse modes and one 3-longitudinal mode with one-loop curvature masses $\hat M^2_{\rm l,t}$, while the mixing between the 4-longitudinal \hbox{(axial-)}vector mode and the scalar mode involves the respective one-loop curvature masses $\hat M^2_{\rm L}$ and $\hat M^2$, all with appropriate flavor indices.

Taking as an example the $V_{4/5}-S_{5/4}$ mixing, one starts from Eq.~\eqref{Eq:det_SV45_final}, writes the classical curvature masses with flavor indices, instead of physical meson indices, and corrects them with the appropriate self-energy. In terms of physical modes, one has
\bea
\det \mathds{M}^{45}_{\mu\nu}=C^2_{45}\left(K^2-\mathfrak{\hat M}^{2,(S)}_{44}\right) \prod_\textrm{p=l,t}\big(K^2-\hat M^{2,(V)}_{55,{\rm p}}\big)^{d_{\rm p}},\qquad
\eea
where $d_{\rm t}=2 d_{\rm l}=2$, $C^2_{45}=\hat M^{2,(V)}_{55,\rm L}-c^2_{45}$ and 
\bea
\mathfrak{\hat M}^{2,(S)}_{44}= Z^2_{S,44} \hat M^{2,(S)}_{44} \qquad  \textrm{with}\qquad Z^2_{S,44}=\frac{\hat M^{2,(V)}_{55,{\rm L}}}{C^2_{45}},\qquad
\eea
with $\hat M^{2,(S)}_{44} = \hat m^{2,(S)}_{44} + \Pi^{(S)}_{44}(0)$ and $\hat M^{2,(V)}_{55,{\rm p}}=m^{2,(V)}_{55} + \Pi^{(V),55}_{\rm p}(0), \textrm{p=L,t,l}.$

All our previous formulas can be modified similarly: $\hat m^{2,(S/P)}_{aa}\to \hat M^{2,(S/P)}_{aa}$, while $\hat m^{2,(V/A)}_{aa}\to \hat M^{2,(V/A)}_{aa,{\rm L}}$ in the $C^2_{ab/aa}$ and $Z^2$ constants, and $\hat m^{2,(V/A)}_{aa}\to \hat M^{2,(V/A)}_{aa,\textrm{l/t}}$ in the former contribution of the propagating \hbox{(axial-)}vector modes. At $T=0$ one has  $\hat M^{2,(V/A)}_{aa,{\rm t/l/L}}=\hat M^{2,(V/A)}_{aa,{\rm vac}}.$

\section{Numerical results \label{sec:num}}

In this section we put to work the formulas derived so far and present in the isospin symmetric case ($\phi_3=0$) the temperature dependence of the one-loop curvature masses obtained for nonvanishing \hbox{(axial-)}vector Yukawa coupling. In order to achieve this, we minimally extend the parametrization used in Ref.~\cite{Kovacs:2016juc} and solve the model using the field equations derived there in the mean-field approximation (see Eq.~(40) there). In that article the model parameters were determined based on one-loop curvature masses for (pseudo)scalar mesons and tree-level ones for \hbox{(axial-)}vector mesons. A parametrization and solution of the model in the proposed localized Gaussian approximation will be presented elsewhere.

\begin{table}[!b]
  \caption{Parameter values in the EL$\sigma$M for our best fit characterized by $\chi^2/N_{\rm d.o.f.}=12.96/15\approx 0.87,$ where $N_{\rm d.o.f.}=15$ is the number of degrees of freedom. The corresponding pseudocritical temperature determined from the inflection point of $\phi_{\rm N}$ is $T_c\approx175$~MeV. $h_1,h_2,$ and $h_3$ are the parameters of the Lagrangian given in Eq.~(2) of Ref.~\cite{Parganlija:2012fy}. $\tilde\kappa$ is the renormalization scale in the $\overline{\rm MS}$ scheme which appears in Eqs.~\eqref{Eq:V_A_2masses_vac_result} and \eqref{Eq:V_A_eq_masses_vac_result}, while $M_0$ is the scale used in Ref.~\cite{Kovacs:2016juc}. \label{Tab:param} }
\centering
\begin{tabular}[c]{cccc}\hline\hline
Parameter & Value & Parameter & Value \\\hline
$\phi_{N}$ [GeV]    & $0.1427$       & $g_{1}$ & $5.9252$\\
$\phi_{S}$ [GeV]    & $0.1405$       & $g_{2}$ & $2.0483$\\
$m_{0}^2$ [GeV$^2$] & $-1.0874\e{-2}$ & $h_{1}$ & $35.5174$\\
$m_{1}^2$ [GeV$^2$] & $1.5428\e{-3}$ & $h_{2}$ & $-12.0902$\\
$\lambda_{1}$       & $-2.0423$      & $h_{3}$ & $4.2493$\\
$\lambda_{2}$       & $24.22$      & $g_{S}$ & $4.5726$\\
$c_{1}$ [GeV]       & $1.1607$       & $g_{V}$ & $5.2818$\\
$\delta_{S}$ [GeV$^2$] & $0.2399$    & $\tilde\kappa=M_0$ [GeV]  & $0.3511$\\\hline\hline
\end{tabular}
\end{table}

Including the Yukawa coupling $g_V$ among the fitting parameters, we determined the EL$\sigma$M parameters using the $\chi^2$ minimization described in Ref.~\cite{Kovacs:2016juc}. We used the same physical quantities as in that article, but replaced the tree-level \hbox{(axial-)vector} curvature mass formulas with the vacuum one-loop level ones. The renormalization scale was fixed to the value used in Ref.~\cite{Kovacs:2016juc}, while for the Polyakov-loop potential we used the parameters given in Table~IV and Fig.~1 of that article. The parameters corresponding to the lowest $\chi^2$ value were found from a fit started in $10^5$ random initial points of the 15-dimensional parameter space, representing the parameters of the EL$\sigma$M Lagrangian. Their values are given in Table~\ref{Tab:param} and can be compared to those appearing in Table~IV of Ref.~\cite{Kovacs:2016juc}. Both parameter sets are compatible with the constraints among $m_0^2,\lambda_1,$ and $\lambda_2$ required by the spontaneous symmetry breaking, which were derived in Ch.~44.13 of Ref.~\cite{Zinn-Justin} from the classical potential.

\begin{figure}[!t]
\hspace*{-0.3cm}
\includegraphics[width=0.5\textwidth]{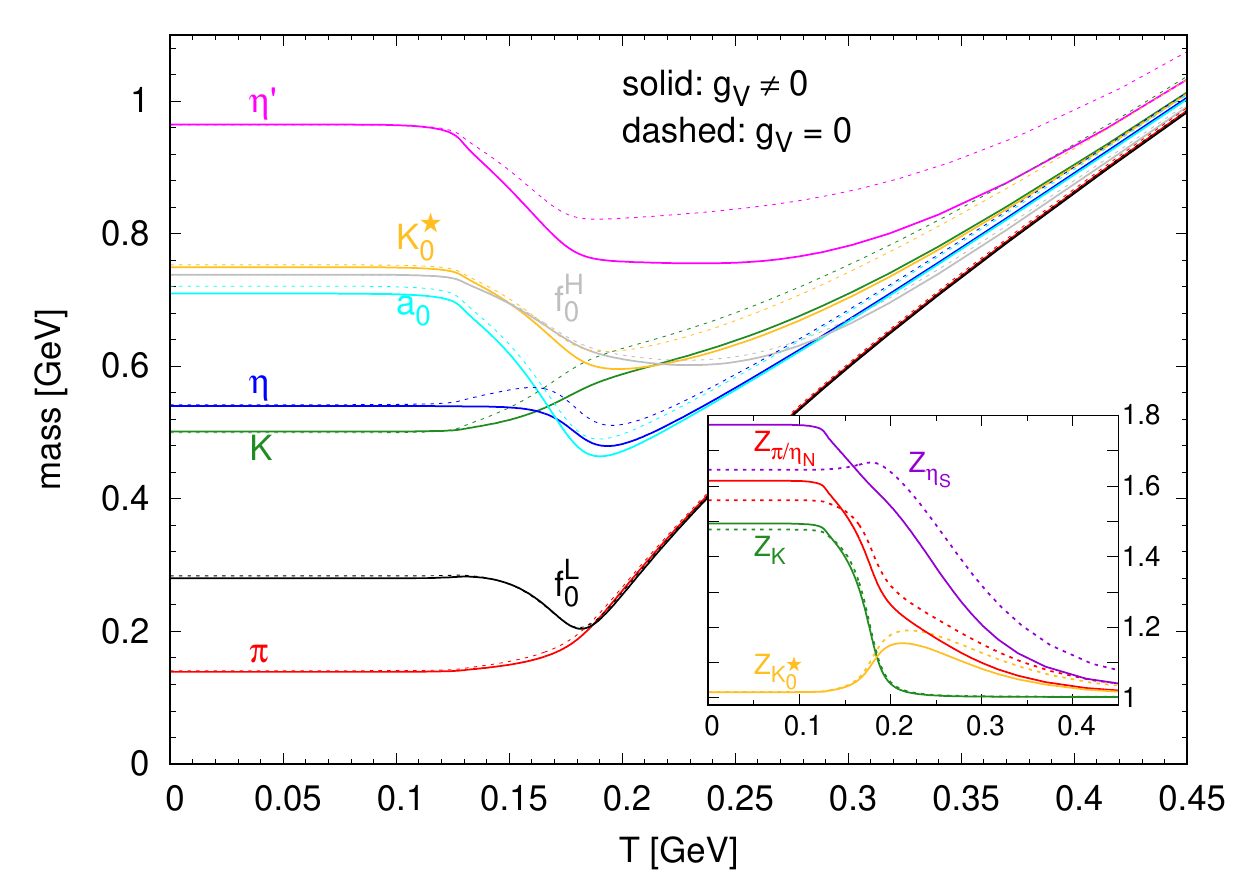}
\caption{Temperature dependence of the (pseudo)scalar one-loop curvature masses obtained in the isospin symmetric case with a parametrization of the model that includes fermionic corrections also in the \hbox{(axial-)}vector masses (solid lines), compared to the case when the parametrization uses tree-level \hbox{(axial-)}vector masses (dashed curves, taken from Ref.~\cite{Kovacs:2016juc}). The underlying field equations are in both cases the mean-field ones given in Eq.~(40) of Ref.~\cite{Kovacs:2016juc}. $f_0^{\rm H/L}$ denotes the eigenstates of the scalar mixing sector with higher/lower mass, respectively. The inset shows the $T$ dependence of the wave-function renormalziation constants.\label{Fig:PS_m-T}}
\end{figure}

\begin{figure}[!t]
\hspace*{-0.3cm}
\includegraphics[width=0.5\textwidth]{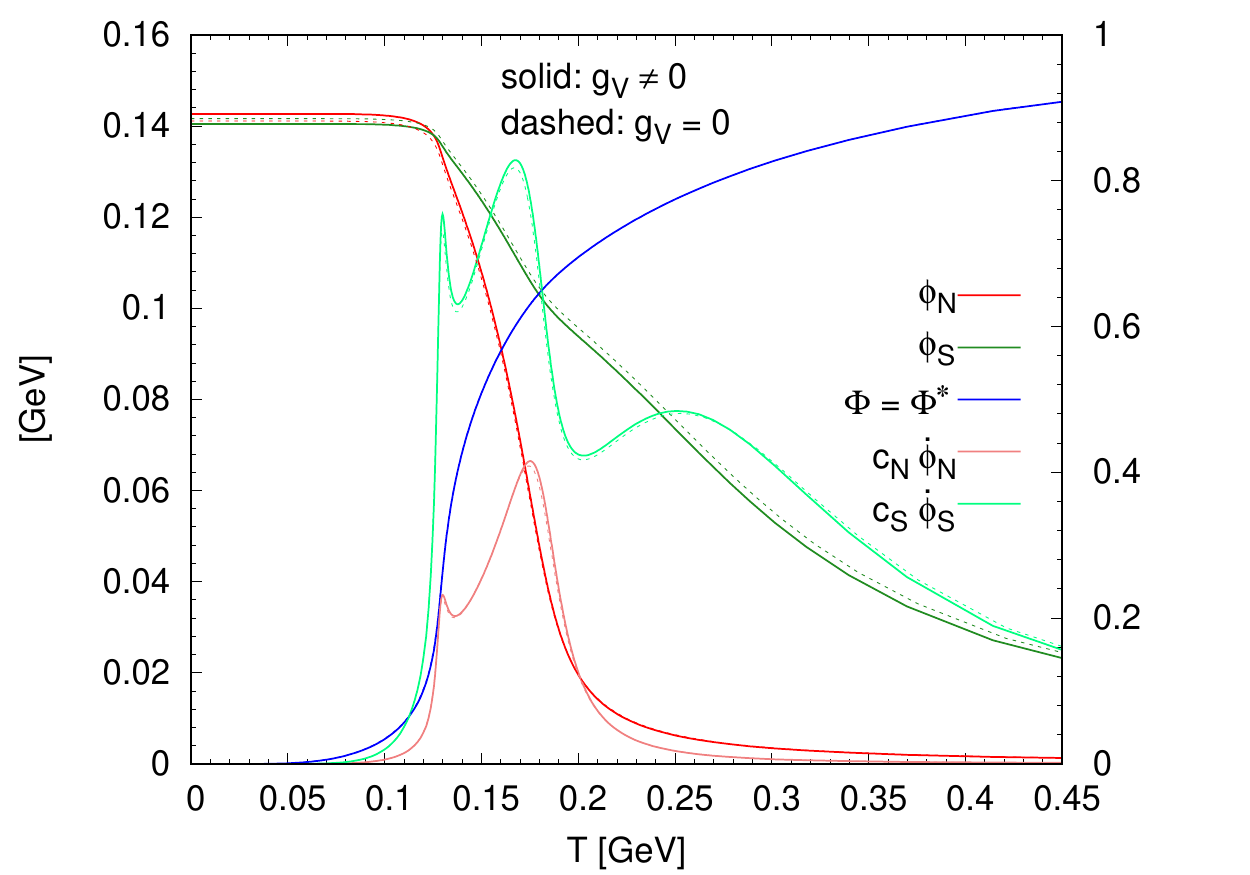}
\caption{Temperature dependence of the scalar condensates and their derivatives and of the Polyakov-loop expectation values compared to the case shown also in Fig.~1 of \cite{Kovacs:2016juc}, where the vector Yukawa coupling is zero, $g_V=0$. The values of the normalization constants used for the sake of presentations are $c_{\rm N} = -0.028$ and $c_{\rm S} = -0.18$, while the position of the global maximum of $\dot \phi_{\rm N}=d\phi_{\rm N}/d T$ gives $T_c=175.238$~MeV and $T_c=174.566$~MeV for the solid and dashed curves, respectively.}
\label{Fig:phiN}
\end{figure}

In Fig.~\ref{Fig:PS_m-T} we compare the $T$ dependence of the (pseudo)scalar masses obtained with a parametrization that takes into account the one-loop contribution of the quarks in the vacuum masses of all the mesons ($g_V\ne0$) to the previous result of Ref.~\cite{Kovacs:2016juc} ($g_V=0$). In the inset we plot the wave-function renormalization constants that correspond to the two cases and are computed with the formulas of Secs.~\ref{ss:GA-mix} and \ref{ss:cl-mix}, respectively. Given that the field equations are the same in both cases and that the parameter values are not much different, we see similar behaviors as the temperature increases. The temperature evolution of the scalar condensates and of the Polyakov-loop expectation values is almost identical to that shown in Fig.~1 of Ref.~\cite{Kovacs:2016juc}, as can be explicitly seen here in Fig.~\ref{Fig:phiN}.

The mass of the pseudoscalars is more affected by the change in the parametrization than the mass of the scalars, especially around the pseudocritical temperature $T_c$ and above it. This is expected because all the pseudoscalar mesons mixes with an axial-vector meson with matching quantum numbers, while from the vector mesons only the mass of $K^\star$ is directly affected by the mixing with the scalar meson $K_0^\star$. Interestingly, the decrease of the $\eta$ and $\eta'$ masses around $T_c$ is more prominent for the parametrization with $g_V\ne0.$ For both parametrizations the $a_0$ meson becomes degenerate with the $\eta$ meson at large $T$. Such a pattern was observed also within the FRG formalism, but only when one goes beyond the local potential approximation \cite{Rennecke:2016tkm}. We also mention that if the model is solved at nonzero temperatures with unchanged parameter values but with all the $Z$ factors set to 1, then $a_0$ degenerates with $\eta'$.

The drop of the $\eta'$ mass around $T_c$ seen in Fig.~\ref{Fig:PS_m-T}, which is observed experimentally in \cite{Vertesi:2009wf}, is accompanied in our case by a drop of the $\eta$ mass. This behavior is related only to the decrease of the $\phi_{N,S}$ condensates, as in \cite{Rennecke:2016tkm}, and not to the restoration of the $U(1)_{A}$ symmetry, which in our case would require a temperature-dependent 't Hooft coupling $c_1.$ The effect of such a coefficient that decreases exponentially with $T^2$ was studied in \cite{Rai:2018ufz} within the $(2+1)$-flavor Polyakov-loop quark–meson model. In \cite{Fejos:2015xca} mesonic fluctuations were incorporated into the axial anomaly in the $N_f=2+1$ flavor linear sigma model using the FRG method in the local potential approximation. The chiral-condensate-dependent anomaly coefficient is subject to its own flow equation, and it was shown that under certain circumstances the thermal evolution of the condensate could induce a reduction of the axial anomaly. However, a careful parametrization of the model done later in \cite{Fejos:2016hbp} showed that the anomaly actually increases around $T_c$. While in that paper $m_\eta$ increases monotonically with the temperature, $m_{\eta'}$ has a nonmonotonic thermal evolution, showing a slight decrease above $T_c,$ before becoming equal with $m_{a_0}$ at high $T$. A direct link between the restoration of $U(1)_A$ symmetry and the drop in the $m_{\eta'}(T)$, without a drop in $m_\eta(T)$, was reported in \cite{Horvatic:2018ztu}. A recent model-independent analysis done in \cite{Nicola:2018vug} suggests that the axial symmetry is restored when the chiral partners become degenerate. 

\begin{figure}[!t]
\hspace*{-0.3cm}  
\includegraphics[width=0.5\textwidth]{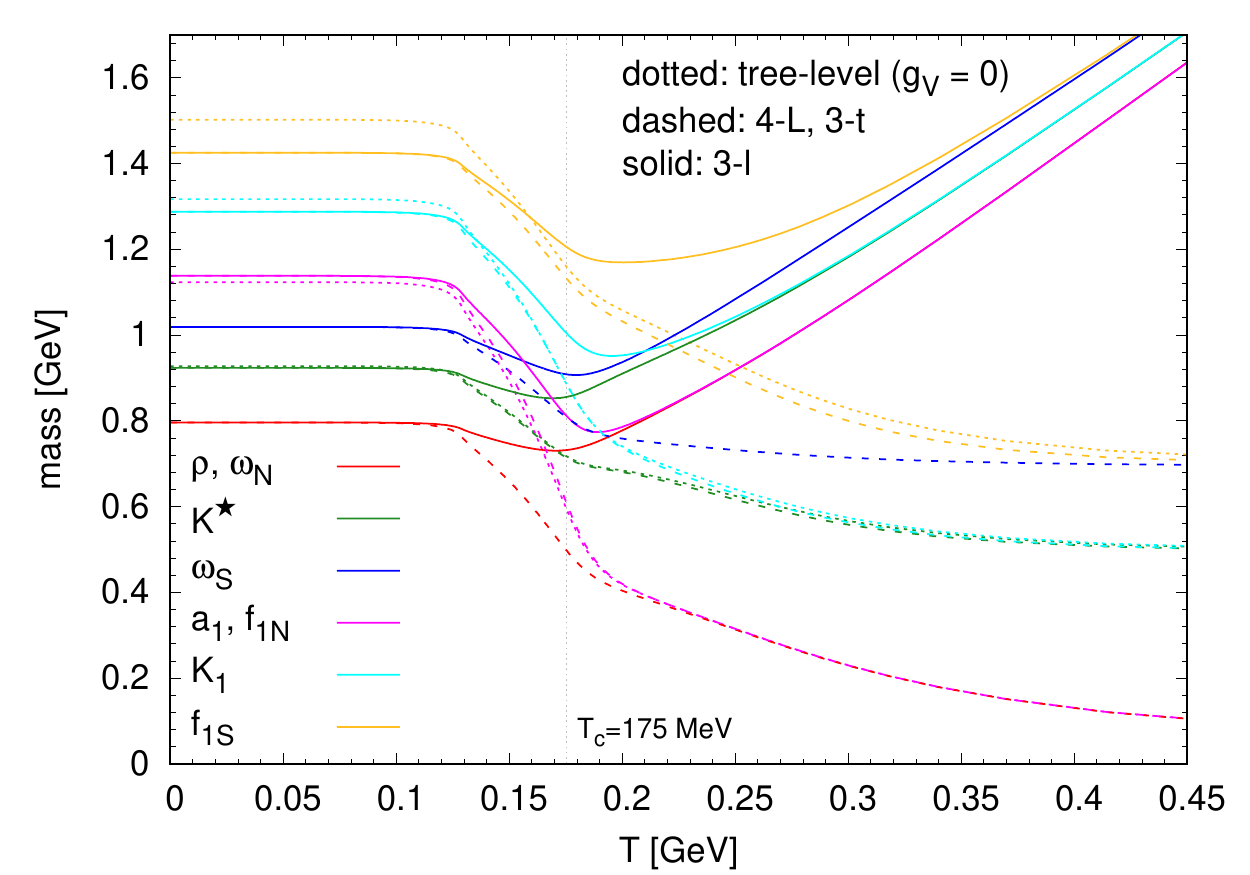}
\caption{Temperature dependence of the \hbox{(axial-)vector} one-loop curvature mass of the 4-longitudinal, 3-transverse/3-longitudinal modes (solid and dashed lines) compared to the tree-level ones (dotted lines). The masses of the 3-transverse and the 4-longitudinal modes are the same, and for the $\rho$ and $\omega_{\rm N,S}$ vector mesons they also coincide with the tree-level mass (not shown in these cases). \label{Fig:AV_m-T}}
\end{figure}

In Fig.~\ref{Fig:AV_m-T} we show the temperature dependence of the one-loop curvature mass of various \hbox{(axial-)}vector modes. In the case of $\rho$ and $\omega$ vector mesons only the mass of the 3-longitudinal mode acquires fermionic correction, and the mass of the other modes remains the tree-level one. In the case of all \hbox{(axial-)}vector mesons this is the mode whose mass increases with increasing temperature deep in the symmetric phase, similarly to the mass of the (pseudo)scalar mesons. Compared to the $N_f=2$ version of the model studied with FRG techniques in \cite{Eser:2015pka,Jung:2016yxl,Jung:2019nnr}, where all the chiral partners degenerate basically at the same temperature, the light vector and axial-vector chiral partners $\rho$ and $a_1$ degenerate at slightly higher temperatures than the (pseudo)scalar ones, $f_0^{\rm L}$ and $\pi$. The chiral partners $K^\star$ and $K_1$ having both light and strange quark content degenerates at a higher temperature than those containing only light quarks, as the strange condensate is still large around the temperature where the nonstrange condensate $\phi_{\rm N}$ melts (see Fig.~\ref{Fig:phiN}). The purely strange chiral partners $\omega_{\rm S}$ and $f_{\rm 1S}$ degenerate at even higher temperatures, where $\phi_{\rm S}$ also melts. The degeneracy of the chiral partners is displayed also by the masses of $4$-longitudinal and $3$-transverse modes. The mass gap between the $3$-longitudinal and $3$-transverse modes increases with $T$ as a result of the violation of the Lorentz symmetry.

\section{Conclusions and outlook \label{sec:sum}}

We investigated the one-loop fermionic contribution to the curvature masses of (pseudo)scalar and \hbox{(axial-)}vector mesons in the framework of a $U(3)_L\times U(3)_R$ linear sigma model with a Yukawa type interaction between mesons and constituent quarks. These corrections were calculated by evaluating the self-energy of the mesons at zero external momentum. It was showed explicitly that this is equivalent to the direct calculation of the second field derivative of the fermionic functional determinant. The one-loop curvature masses of the \hbox{(pseudo)}scalars agree with those derived in Ref.~\cite{Schaefer:2008hk} with an alternative method that uses fluctuation-dependent quark masses. We pointed out that this alternative method cannot be used for the \hbox{(axial-)}vector mesons due to the presence of the momentum-dependent Lorentz scalars $V^{\mu}Q_{\mu}$ and $A^{\mu}Q_{\mu}$ in the fermion determinant.

The renormalization of the curvature masses was discussed in detail. The divergencelessness of the vector current, which occurs on a specific scalar background for certain flavor indices (\emph{e.g.} for $a\ne4-7$ in the isospin symmetric case), has the consequence that the corresponding vector self-energy is 4-transverse and vanishes at zero momentum. To comply with this property a suitable regularization scheme is needed. To keep the discussion uniform, dimensional regularization was used in the renormalization of both the vector and the \hbox{(axial-)}vector self-energies for all flavor indices. Additionally, the renormalization revealed that a chiral-invariant term appeared twice in the EL$\sigma$M Lagrangian \cite{Parganlija:2010fz, Parganlija:2012fy}. This can be cured with the appropriate redefinition of some couplings.

The occurrence of the $S-V$ and $P-A$ mixing already showed the importance of the mode decomposition of the \hbox{(axial-)}vector self-energy, which was investigated in detail at both $T=0$ and $T\ne0$, as the 4-longitudinal mode of the \hbox{(axial-)}vectors mixes with the (pseudo)scalars. As a result, in the case of the Gaussian approximation, the one-loop curvature mass of the (pseudo)scalar mesons is modified by a wave-function renormalization constant determined in terms of the one-loop curvature mass of the 4-longitudinal \hbox{(axial-)}vector mode. In a simpler approximation we recovered the already known versions of these constants appearing in \cite{Parganlija:2012fy}.
    
The vacuum parametrization of the model was redone based on curvature masses that include one-loop fermionic contributions for all the mesons. The temperature dependence of all these masses was investigated. The \hbox{(axial-)}vector tensor splits up into 3-transverse modes (which turns out to have the same contribution as the 4-longitudinal one) and a 3-longitudinal mode. In the isospin symmetric case the mass of 3-transverse modes of the vector mesons $\rho$, $\omega$ (or $\omega_{\rm N}$) and $\phi$ (or $\omega_{\rm S}$) coincides with the corresponding tree-level mass, while for the other particles the mass of the 3-transverse modes is slightly different from the tree-level mass. For all \hbox{(axial-)vector} particles the mass of the 3-longitudinal mode significantly deviates from the tree-level one. It increases with increasing temperature, similarly to the (pseudo)scalar curvature mass, while the mass of the 3-transverse components decreases with increasing temperature. The particle masses of the two modes become degenerate separately as the chiral symmetry restores with increasing temperature and the mass gap increases between them as a reflection of Lorentz symmetry violation.

As a side benefit of the new parametrization of the model, the value of the vector meson Yukawa coupling $g_V$ was determined. This value influences the equation of state used to describe properties of the compact star, where it has a prominent role in determining the maximal value of the compact star mass of the $M$-$R$ curves (see \emph{e.g.} Ref.~\cite{Masuda:2012ed}).

The curvature masses of the various \hbox{(axial-)}vector modes determined here can be used not only in the localized Gaussian approximation proposed in Sec.~\ref{sec:GA}, but also in the localized version of the two-particle irreducible formalism in which in \cite{Struber:2007bm} the gauged version of the purely mesonic model was solved at two-loop level for $N_f=2$. In the latter context the mode decomposition of self-energy presented here would allow for an improved approximation, as there the complexity of the numerical problem was reduced by using even at finite temperature a curvature mass tensor of a vacuum form, that is, proportional to $g_{\mu\nu}$.

\begin{acknowledgments}
We thank Gy.~Wolf for including $g_V$ among the parameters of the code used in Ref.~\cite{Kovacs:2016juc} for the parametrization of the model. Zs.~Sz is thankful to Gergely Fej\H{o}s and Matteo Giordano for valuable discussions on the subject. The research was supported by the Hungarian National Research, Development and Innovation Fund under Project No.~FK 131982. P.~K. also acknowledges support by the J\'anos Bolyai Research Scholarship of the Hungarian Academy of Sciences.
\end{acknowledgments}

\appendix

\section{The components of $\Pi_{\rm mat}^{(V),\mu\nu}$ for $N_f=1$ \label{app:check_comp}}

It is instructive to examine the \emph{matter} part of the vector self-energy at vanishing momentum \eqref{Eq:polarisation_int}. The $0i$ matrix element vanishes by symmetric integration, while the temporal and spatial components are different due to the breaking of the Lorentz symmetry by the heat bath.

Using $k_0^2=\k^2+m_f^2+G_f^{-1}$ for the 00 component and
\be
\int d^3 k \,k_i k_j f(\k^2)=\frac{\delta_{ij}}{3}\int d^3 k\, \k^2 f(\k^2)
\label{Eq:rot_inv_3d}
\ee
for the $ij$ component, one obtains (${\cal T}_f^{(1)}\equiv{\cal T}^{(1)}(m_f)$)
\bea
\Pi^{(V),00}_\textrm{mat}&=&4 g_V^2\bigg[{\cal T}_f^{(1)}+2 m_f^2\frac{d {\cal T}_f^{(1)}}{d m_f^2}+2\frac{d{\cal U}^{(1)}(m_f)}{d m_f^2}\bigg],\qquad  \label{Eq:dhatM2_V_00} \\
\Pi^{(V),ij}_\textrm{mat}&=&4 g_V^2\delta^{ij}\bigg[{\cal T}^{(1)}_f+\frac{2}{3}\frac{d{\cal U}^{(1)}(m_f)}{d m_f^2}\bigg]. \label{Eq:dhatM2_V_ij}
\eea
Here we used the \emph{matter} part of the integral ${\cal U}(m_f)=i\int_K k^2 G_f(K)$, which -- having only an extra factor of $k^2\equiv\k^2$ compared to the tadpole defined in Eq.~\eqref{Eq:tad} -- is given (see Eqs.~\eqref{Eq:ITF} and \eqref{Eq:tad_FD_T}) by
\be
{\cal U}^{(1)}(m_f)=-\frac{1}{4\pi^2}\int_0^\infty\textrm{d} k\frac{k^4}{E_f(k)}\big[f_f^+(k)+f_f^-(k)\big].
\label{Eq:U_finite_T}   
\ee   
With a partial integration, as in Eq.~\eqref{Eq:Bub_T}, one obtains
\be
\frac{d{\cal U}^{(1)}(m_f)}{d m_f^2}=-\frac{3}{2} {\cal T}^{(1)}(m_f).
\label{Eq:rel_dUdm2-T}
\ee

As a result of Eq.~\eqref{Eq:rel_dUdm2-T} we see from Eq.~\eqref{Eq:dhatM2_V_ij} that $\Pi^{(V),11}_{\rm mat}=0,$ and therefore $\tr_{\rm L}\Pi^{(V)}_\textrm{mat}=\Pi^{(V),00}_{\rm mat}-3 \Pi^{(V),11}_{\rm mat}=\Pi^{(V),00}_{\rm mat},$ and thus from Eq.~\eqref{Eq:dhatM2_V_00} one obtains
\be
\tr_{\rm L}\Pi^{(V)}_\textrm{mat}=\Pi^{(V),00}_\textrm{mat}=-8 g_V^2 \bigg[1 - m_f^2\frac{d}{d m_f^2} \bigg]{\cal T}_f^{(1)}.
\ee
This expression agrees (up to a convention related sign) to that obtained from Eq.~(5.51) of Ref.~\cite{Kapusta-Gale} by taking there $\k\to 0$ at $k_0=0$ (the limits do not commute).

\section{Vacuum property of $\Pi^{(V)\mu\nu}_{ab}(Q)$ for certain flavor indices \label{app:PiV_prop}}

The vector fields $V_\mu^a, a=0,\dots, 8$ couple in the $N_f=2+1$ version of the Lagrangian \eqref{Eq:L_f} to the vector current\footnote{Note that we call $J_a^{(V)\mu}$ the vector current by abuse of terminology and by analogy with QCD, as in the context of the linear sigma model the true vector current also contains mesonic fields.} $J_a^{(V)\mu}=\bar\psi \gamma^\mu T_a\psi$, where $T_a=\lambda_a/2$. Using the Euler-Lagrange equations one finds
\be
\partial_\mu J_a^{(V)\mu} = i\bar\psi \comm{M}{T_a}\psi,
\label{Eq:current_div}
\ee
where
\be
M=\left[g_S(S^a-i\gamma_5 P^a) +g_V \gamma^\mu \big(V^a_\mu + \gamma_5 A^a_\mu\big) \right]T_a.
\label{Eq:M_def}
\ee
The divergence of the current \eqref{Eq:current_div} has the same form as in QCD \cite{Scherer:2012xha}, where $M$ is the current quark matrix.

Interestingly, when evaluated on the background of the mesonic fields, that is, with
\be
\bar M=\frac{g_S}{2}\sum_{b=0,3,8}\phi_b\lambda_b=
\frac{g_S}{2}\sum_{b=3,\textrm{N,S}}\phi_b\lambda_b,
\label{Eq:bar_M_def}
\ee
the divergence of the vector current vanishes for certain flavor indices, namely
\be
\partial_\mu J_a^{(V)\mu}|_{\bar M}=\begin{cases}
0, \ \ \textrm{for}\ a=3,{\rm N,S}\ \ \textrm{when}\  \phi_3\ne 0,\\
0, \ \ \textrm{for}\ a=1,2,3,{\rm N,S}\ \  \textrm{when}\  \phi_3 = 0,
\end{cases}
\label{Eq:zero_div}
\ee
as one can check using the values of the antisymmetric structure constant.

We show below that at quantum level the divergencelessness of the current has the consequence that the \emph{vacuum} part of the one-loop self-energy defined in Eq.~\eqref{Eq:SE_Nf3} satisfies
\be
Q_\mu \Pi_{ab,{\rm vac}}^{(V)\mu\nu}(Q)=0 \quad \Longrightarrow \quad \Pi_{ab, {\rm vac}}^{(V)\mu\nu}(Q=0)=0,
\label{Eq:relation_for_Pi_V}
\ee
in the cases listed in Eq.~\eqref{Eq:zero_div}, where from Table~\ref{Tab:flavor_tr} we know that the self-energy is nonzero only when $b=a$ (for the implication above see p.~233 of Ref.~\cite{Lurie}).

Considering the meson fields as classical external fields, we start by relating the expectation value of the current and its divergence with the fermion propagator matrix (see p.~66 in \cite{Fetter-Walecka})
\begin{subequations}
\bea
\label{Eq:J-S_rel}
\big\langle J_a^{(V)\mu}(x)\big\rangle &=& -\lim_{y\to x}\tr\left(\gamma^\mu T_a \bar {\cal S}(x,y)\right),\\
\big\langle \partial_\mu J_a^{(V)\mu}(x)\big\rangle &=& -\lim_{y\to x}\tr\left(\comm{M(x)}{T_a} \bar {\cal S}(x,y)\right).\ \ 
\label{Eq:dJ-S_rel}
\eea
\end{subequations}
The trace in Eq.~\eqref{Eq:J-S_rel} is to be taken in color, flavor, and Dirac spaces. In the SSB case, when the fields are shifted with their expectation values, the full propagator obeys 
\be
\left(i\slashed{\partial}_x-\bar M - M(x)\right)\bar{\cal S}(x,y) = i \delta^{(4)}(x-y),
\label{Eq:matrix_prop_S}
\ee
with $M(x)$ given in Eq.~\eqref{Eq:M_def} and $\bar M$ in Eq.~\eqref{Eq:bar_M_def}.

Next, we expand the full propagator about the tree-level propagator introduced in Sec.~\ref{sec:SE_Nf_3}, which obeys $\left(i\slashed{\partial}_x-\bar M\right)\bar{\cal S}_0(x,y) = i \delta^{(4)}(x-y).$ To do so, we write the formal solution of Eq.~\eqref{Eq:matrix_prop_S} as $\bar{\cal S}=i/(A-M)$, where $A=i\bar {\cal S}_0^{-1}$, and use
\bea
\frac{1}{A-M} &=& \frac{1}{A}(A-M+M)\frac{1}{A-M} = \frac{1}{A} +\frac{1}{A}M\frac{1}{A-M} \nonumber \\
&=& \frac{1}{A} +\frac{1}{A}M\frac{1}{A} + \dots\ .
\eea
This gives at one-loop level
\be
\bar{\cal S}(x,y)\simeq \bar{\cal S}_0(x,y) - i\int_z \bar{\cal S}_0(x,z) M(z) \bar{\cal S}_0(z,y).
\label{Eq:S_expansion}
\ee
Taking the derivative of Eq.~\eqref{Eq:J-S_rel} and using Eq.~\eqref{Eq:S_expansion} we obtain
\be
\partial_\mu \big\langle J_a^{(V)\mu}(x)\big\rangle
\simeq i\partial_x^\mu\tr \int_y \gamma_\mu \frac{\lambda_a}{2} \bar {\cal S}_0(x,y) M(y) \bar{\cal S}_0(y,x),
\label{Eq:CC_1-loop}
\ee
where the contribution of $\bar{\cal S}_0(x,y)$ from Eq.~\eqref{Eq:S_expansion} vanishes due to translational invariance.

It would be tempting to say that the left-hand side of Eq.~\eqref{Eq:CC_1-loop} vanishes as result of Eq.~\eqref{Eq:zero_div}, but the usual proof using the invariance of the functional integral with respect to the vector $U(3)_V$ transformation does not go through because we neglected the mesonic fields in Eq.~\eqref{Eq:current_div} and the current vanishes only on a specific scalar background. What is easy to prove however, is that Eq.~\eqref{Eq:dJ-S_rel} vanishes at linear order in $M(x)$, \emph{i.e.} the order at which Eq.~\eqref{Eq:CC_1-loop} was derived. Indeed, using the first term in Eq.~\eqref{Eq:S_expansion}, one has in the cases listed in Eq.~\eqref{Eq:zero_div}
\bea
\partial_\mu \big\langle J_a^{(V)\mu}(x)\big\rangle &=& \big\langle \partial_\mu J_a^{(V)\mu}(x)\big\rangle\nonumber\\
&\simeq& -\tr\left(\comm{M(x)}{T_a} \bar {\cal S}_0(x,x)\right) = 0,
\label{Eq:zero_div_Qlevel}
\eea
because $\displaystyle \bar {\cal S}_0 = u_c\lambda_c$, $c=0,3,8$, with $u_3=0$ for $\phi_3=0$, and the structure constant is such that $\tr_{\rm F}(\comm{\lambda_b}{\lambda_a}\lambda_c)u_c=4 i f_{bac} u_c=0$ for $c=0,3,8$, $b=0,\dots,8$ and $a$ taking the values given in Eq.~\eqref{Eq:zero_div} (instead of ${\rm N}$ and ${\rm S}$ one can use $0$ and $8$).

Since for the flavor indices\footnote{For other flavor indices the scalar term can also contribute.} of Eq.~\eqref{Eq:zero_div} only the vector term in $M(x)$ contributes in Eq.~\eqref{Eq:CC_1-loop}, we obtain using Eq.~\eqref{Eq:zero_div_Qlevel}
\be
0 = i g_V \partial_x^\mu \int_y \tr\Big[\gamma_\mu \frac{\lambda_a}{2} \bar {\cal S}_0(x,y) \gamma_\nu \frac{\lambda_b}{2} \bar{\cal S}_0(y,x)\Big] V_b^\nu(y). \ \ \ \ 
\label{Eq:Eq:CC_1-loop_final}
\ee
Going to momentum space and using the definition \eqref{Eq:SE_Nf3} of the self-energy, one easily obtains Eq.~\eqref{Eq:relation_for_Pi_V}, which holds in the cases listed in Eq.~\eqref{Eq:zero_div}.

\section{Integrals determining the self-energy for $N_f=2+1$ \label{app:PiCalcDetails}}

Here we calculate the integrals \eqref{Eq:I_X_def_zero_mom} relevant for the expression of the curvature mass.

In the case of a \hbox{(pseudo)}scalar field one has $\Gamma_S=\mathbb{1}$ ($\Gamma_P=\gamma_5$) and the Dirac trace gives ($G_f^{-1}(K)=K^2-m_f^2$)
\be
\tr_{\rm D}\big[\Gamma_X {\cal S}_f\Gamma_X {\cal S}_{f'}\big] = -4(m_f m_{f'}\pm K^2) G_f G_{f'},
\label{Eq:S_P_2masses}
\ee
with a plus sign for the scalar. One adds and subtracts in the numerators produced by partial fraction decomposition the mass squared of the corresponding denominators to obtain
\bea
I^{S/P}(m_f,m_{f'})=\pm \frac{m_f {\cal T}(m_f) \mp m_{f'} {\cal T}(m_f')}{m_f \mp m_{f'}}.
\eea
For equal masses the limit in Eq.~\eqref{Eq:I_X_one_mass} gives (${\cal T}_f\equiv{\cal T}(m_f)$)
\be
I^S(m_f) = \bigg(1 + 2m_f^2\frac{d}{d m_f^2}\bigg) {\cal T}_f, \quad I^P(m_f) = -{\cal T}_f,
\ee
Using Eqs.~\eqref{Eq:SE_Nf3} and \eqref{Eq:I_X_def} and Tables~\ref{Tab:couplings} and \ref{Tab:flavor_tr} one obtains, for example, $\displaystyle \Pi_{11}^{(S/P)}(Q=0)=\mp 2 N_c g_S^2 I^{S/P}(m_u,m_d).$

For \hbox{(axial-)}vector fields we start with the momentum-dependent integral in Eq.~\eqref{Eq:I_X_def}, as this is needed in some relations derived in Appendix~\ref{app:prop_self_energy}. Using $\Gamma_V=\gamma_\mu$ and $\Gamma_A=\gamma_\mu\gamma_5$, the Dirac trace gives ($P=K-Q$)
\bea
&&I^{V/A,\mu\nu}(Q;m_f,m_{f'})=i\int_K G_f(K) G_{f'}(P)\nonumber \\
&&\times \big[(\pm m_f m_{f'}-K\cdot P)g^{\mu\nu} + K^\mu P^\nu + K^\nu P^\mu \big],
\label{Eq:V_A_2masses}
\eea
with the upper sign for the vector. Next we consider separately the \emph{vacuum} and the \emph{matter} parts of this integral.

\paragraph{Vacuum part} In a covariant calculation at $T=0,$ Feynman parametrization and dimensional regularization give ($d=4-2\epsilon$)
\bea
&&I_{\rm vac}^{V/A,\mu\nu}(Q;m_f,m_{f'})=\frac{\Gamma(2-\frac{d}{2})}{(4\pi)^{d/2}} \kappa^{4-d}\int_0^1 d x \Delta^{\frac{d}{2}-2}\nonumber\\
&&\times \big[(M^2(x)\mp m_f m_{f'})g^{\mu\nu}+2x(1-x)Q^2 \textrm{P}_{\rm T}^{\mu\nu}(Q)\big].
\label{Eq:V_A_2masses_vac}
\eea
Here $\kappa$ is the renormalization scale, $\textrm{P}_{\rm T}^{\mu\nu}(Q)$ is the 4-transverse projector \eqref{Eq:PLT_zeroT}, and $\Delta=M^2(x)-x(1-x)Q^2$ with $M^2(x)=(1-x)m_f^2+x m_{f'}^2.$ We see that
\be
\lim_{m_{f'}\to m_f} (M^2(x)\mp m_f m_{f'})=\begin{cases}
0, & \textrm{for}\ V\, (-)\\
2 m_f^2, & \textrm{for}\ A\, (+)
\end{cases},
\ee
and therefore
\be
Q_\mu I_{\rm vac}^{V,\mu\nu}(Q;m_f)=0\quad \textrm{and} \quad I_{\rm vac}^{V,\mu\nu}(m_f)=0.
\label{Eq:IV_relations}
\ee

At vanishing momentum, where
\be
I_{\rm vac}^{V/A,\mu\nu}(Q=0;m_1,m_2)=g^{\mu\nu}I_{\rm vac}^{V/A}(m_1,m_2),
\ee
we split the integral into divergent and finite parts: 
\be
\begin{aligned}
&I_{\rm vac}^{V/A}(m_1,m_2)=I_{\rm vac, D}^{V/A}(m_1,m_2) + I_{\rm vac,F}^{V/A}(m_1,m_2),\\
  &I_{\rm vac,D}^{V/A}(m_1,m_2) = \frac{1}{32\pi^2\epsilon}\, (m_1\mp m_2)^2\, ,\\
  &\begin{aligned}I_{\rm vac,F}^{V/A}(m_1,m_2) =& \frac{1}{64\pi^2}\bigg[m^2_1 + m^2_2\mp 4 m_1 m_2\\
     & - 4\frac{f_\mp(m_1,m_2)-f_\mp(m_2,m_1)}{m^2_1 - m^2_2}\bigg],
   \end{aligned}
\end{aligned}
\label{Eq:V_A_2masses_vac_result}
\ee
where $f_\mp(x,y)=x^3(x\mp 2y)\ln(x/\tilde\kappa)$ and the divergence was given in the $\overline{\rm MS}$ scheme ($\tilde\kappa^2=4\pi e^{-\gamma}\kappa^2$). To obtain the finite part, the $x$ integral in Eq.~\eqref{Eq:V_A_2masses_vac} was evaluated to $\cal{O}(\epsilon),$ as the prefactor contains $1/\epsilon.$

For equal masses one has, in accordance with Eq.~\eqref{Eq:IV_relations},
\bea
&&I_{\rm vac,F}^V(m)=I_{\rm vac,D}^V(m)=0.\nonumber\\
&&I_{\rm vac,F}^A(m)=-\frac{m^2}{8\pi^2}\ln\frac{m^2}{\tilde\kappa^2},\quad I_{\rm vac,D}^A(m)=\frac{m^2}{8\pi^2\epsilon}.\quad
\label{Eq:V_A_eq_masses_vac_result}
\eea

It is easy to see that $I^{V/A}_{\rm vac}(m_1,m_2)$ can be given in terms of tadpole integrals. Indeed, setting $Q=0$ in Eq.~\eqref{Eq:V_A_2masses} one uses the identity
\be
\int d^d K \,K_\mu K_\nu f(K^2)=\frac{g_{\mu\nu}}{d}\int d^d K\, K^2 f(K^2)
\label{Eq:rot_inv}
\ee
and partial fractioning to obtain
\bea
&&I^{V/A}_{\rm vac}(m_1,m_2) = \frac{1}{d} \, \frac{m_1 {\cal T}_\epsilon^{(0)}(m_1) \mp m_2 {\cal T}_\epsilon^{(0)}(m_2)}{m_1 \mp m_2}\nonumber\\
&&\qquad -\frac{d-1}{d}\, \frac{m_1 {\cal T}_\epsilon^{(0)}(m_1) \pm m_2 {\cal T}_\epsilon^{(0)}(m_2) }{m_1 \pm m_2},
\label{Eq:I_VA_tad_reduced}
\eea
and for equal masses 
\be
I^{V/A}_{\rm vac}(m)=-\frac{d-2}{d}\Big[1 + \frac{2z_{V/A}}{d-2} m^2 \frac{d}{d m^2}\Big]{\cal T}_\epsilon^{(0)}(m),
\label{Eq:I_VA_tad_reduced_eq_masses}
\ee
with $z_V=-1$ and $z_A=d-1.$ Here we defined
\be
{\cal T}^{(0)}_\epsilon(m)=\kappa^{4-d}\int\frac{d^d K}{(2\pi)^d}\frac{i}{K^2-m^2}=\frac{\Gamma(2-\frac{d}{2})}{(4\pi)^{d/2}} \frac{\kappa^{4-d}}{(m^2)^{1-\frac{d}{2}}},
\label{Eq:Tvac_dimreg}
\ee
which can be split into divergent and finite parts as
\bea
\label{Eq:Tvac_dimreg_split}
&&{\cal T}^{(0)}_\epsilon(m)={\cal T}^{(0)}_{\epsilon, \rm D}(m) + {\cal T}^{(0)}_{\epsilon, \rm F}(m),\\\nonumber
&&{\cal T}^{(0)}_{\epsilon, \rm D}(m)=\frac{m^2}{16\pi^2}\left[\frac{1}{\epsilon}-1\right],\quad 
{\cal T}^{(0)}_{\epsilon, \rm F}(m)=\frac{m^2}{16\pi^2}\ln\frac{m^2}{\tilde\kappa^2}.
\eea
Using Eq.~\eqref{Eq:Tvac_dimreg} or \eqref{Eq:Tvac_dimreg_split} in Eqs.~\eqref{Eq:I_VA_tad_reduced} and \eqref{Eq:I_VA_tad_reduced_eq_masses}, one recovers Eqs.~\eqref{Eq:V_A_2masses_vac_result} and \eqref{Eq:V_A_eq_masses_vac_result}.

\paragraph{Matter part} We evaluate the integral in Eq.~\eqref{Eq:V_A_2masses} for $Q=0.$ The $0i$ component of the integral vanishes, while for the 00 component we use $k_0^2=K^2+\k^2$ to write
\bea
&&I^{V/A,00}_{\rm mat}(m_f,m_{f'}) = i\int_K\frac{K^2+2\k^2\pm m_f m_{f'}}{(K^2-m_f^2)(K^2-m_{f'}^2)}\bigg|_\textrm{mat}\nonumber\quad\\
&&\qquad\qquad=\frac{m_f {\cal T}^{(1)}_f \mp m_{f'} {\cal T}^{(1)}_{f'}}{m_f \mp m_{f'}} + 2 \frac{{\cal U}^{(1)}_f-{\cal U}^{(1)}_{f'}}{m_f^2-m_{f'}^2}.
\label{Eq:IAV_00_mat}
\eea
Here we used partial fractioning, as described below Eq.~\eqref{Eq:S_P_2masses}, and then Eq.~\eqref{Eq:ITF} to write the result in terms of the integrals given in Eqs.~\eqref{Eq:tad_FD_T} and \eqref{Eq:U_finite_T}.

The $ij$ component of the integral is proportional with $\delta_{ij},$ and thus it is enough to give the 11 component. Setting $Q=0$ in Eq.~\eqref{Eq:V_A_2masses}, we use Eq.~\eqref{Eq:rot_inv_3d}, followed by partial fractioning, as in Eq.~\eqref{Eq:IAV_00_mat}, to obtain
\be
I^{V/A,11}_{\rm mat}(m_f,m_{f'})=\frac{m_f {\cal T}^{(1)}_f \pm m_{f'} {\cal T}^{(1)}_{f'}}{m_f \pm m_{f'}} + \frac{2}{3} \frac{{\cal U}^{(1)}_f-{\cal U}^{(1)}_{f'}}{m_f^2-m_{f'}^2}.
\label{Eq:IAV_11_mat}
\ee
Using Eq.~\eqref{Eq:rel_dUdm2-T}, the equal mass limit of Eqs.~\eqref{Eq:IAV_00_mat} and \eqref{Eq:IAV_11_mat} is
\bea
&&I^{V, 00}_{\rm mat}(m_f)=-2\big[{\cal T}_f^{(1)}+m_f^2{\cal B}_f^{(1)}\big], \quad I^{V, 11}_{\rm mat}(m_f)=0,\qquad \nonumber \\
&&I^{A, 00}_{\rm mat}(m_f)=-2{\cal T}_f^{(1)}, \quad I^{A, 11}_{\rm mat}(m_f)=-2 m_f^2 {\cal B}_f^{(1)},
\eea
with ${\cal B}_f^{(1)}={\cal B}^{(1)}(m_f)$ being the \emph{matter} part of the Euclidian bubble integral at zero momentum given in Eq.~\eqref{Eq:Bub_T}.

\section{Brute force calculation of the curvature mass in the $N_f=2+1$ case \label{app:brute_force}}

After shifting the scalar fields with their expectation values, the integration over the fermionic fields in the Euclidean partition function gives (see Ch.~2.5 of \cite{Kapusta-Gale}) the expression on the right-hand side of Eq.~\eqref{Eq:U_f} with
\bea
&&{\cal S}^{-1}_{{\rm E},f}(K;\xi) = -\textrm{diag}(k^0_u,k^0_d,k^0_s)\otimes \mathbb{1}_{4\times 4}+\mathbb{1}_{3\times 3}\otimes\gamma^0 \vec{\gamma}\cdot\vec{k}\nonumber\\
&&+ \textrm{diag}(m_u,m_d,m_s)\otimes\gamma^0 + g_S\big[S\otimes\gamma^0 -i P\otimes \gamma^0\gamma^5\big]\nonumber\\
&&+ g_V\big[V \otimes \gamma^0\gamma^\mu +A \otimes \gamma^0\gamma^\mu \gamma_5\big]\,,
\label{Eq:matrix_Nf3}
\eea
where $\xi=\{S_a,P_a,V_a^\mu,A_a^\mu|a=1,\dots,7,\textrm{N},\textrm{S}\}$ denotes the set of fields contained in the nonets, $\otimes$ is the Kronecker product, $m_f,$ $f=u,d,s,$ is the constituent quark mass given in Eq.~\eqref{Eq:m_f_tree}, while $k^0_f=i\nu_n+\mu_f$, with $\nu_n$ the Matsubara frequency and $\mu_f$ the chemical potential.

We calculate the determinant of ${\cal S}^{-1}_{{\rm E},f}(K;\xi)$ with the symbolic program MAPLE keeping only those \hbox{(pseudo)}scalar or \hbox{(axial-)}vector fields which are used for differentiation in Eq.~\eqref{Eq:dhatM2_SPVA} and setting to zero the remaining set of fields, denoted as $\xi'=\xi\setminus\{X_a\}.$ This simplified determinant is evaluated in Dirac and flavor spaces and denoted as $D^{(X_a)}:=\det\big [{\cal S}^{-1}_{{\rm E},f}(K;\xi)\big]|_{\xi'=0}$. We found that it can have two forms: for the mixing sector involving the fields $X_3,$ $X_{\rm N},$ and $X_{\rm S}$ the three quark sectors completely factorize, while for fields with other flavor indices there is a mixing between two quark sectors.

The contribution of the scalar ($X=S$) and pseudoscalar ($X=P$) mixing sectors can be written with  $X_\pm=(X_{\rm N}\pm X_3)/\sqrt{2}$ in the following factorized form
\be
D^{(X_{\textrm{N,S},3})}=\prod_{i=\pm,{\rm S}} \left[\frac{g_S^2}{2} X_i^2 + c^{(X)} g_S m_i X_i -G_i^{-1}(K)\right]^2,
\label{Eq:simple_det_mix_SP}
\ee
where $c^{(S)}=\sqrt{2},$ $c^{(P)}=0$ and $G_i^{-1}=K^2-m_i^2$ with $m_+=m_u,$ $m_-=m_d$ and $m_{\rm S}=m_s.$ Inside the square brackets there is no summation over $i$.

For the $V^\mu_\pm=(V_{\rm N}^\mu\pm V_3^\mu)/\sqrt{2}$ vector fields one has
\be
D^{(V_{\textrm{N,S},3})}=\prod_{i=\pm,{\rm S}} \left[\frac{g_V^2}{2}V_i^2 + \sqrt{2} g_V V_i\cdot K + G_i^{-1}(K)\right]^2,
\ee
where $V_i^2=V_i\cdot V_i=V_i^\mu V_{i,\mu},$ while for the axial-vectors $A^\mu_\pm=(A_{\rm N}^\mu\pm A_3^\mu)/\sqrt{2}$ the factorized form is
\begin{align}
  D^{(A_{\textrm{N,S},3})}=&\prod_{i=\pm,{\rm S}} \bigg[\frac{g_V^4}{4}\big(A_i^2\big)^2 + G_i^{-2}(K) \nonumber\\ &
    + g_V^2 \Big( A_i^2\big(K^2+m_i^2\big) - 2 (A_i\cdot K)^2\Big) \bigg].
\end{align}

When there is no complete factorization of quark sectors, the contribution for scalar ($X=S,$ upper sign) and pseudoscalar ($X=P$, lower sign) is
\begin{align}
D^{(X_a)}=&\bigg[\frac{g_S^4}{16}X_a^4 - \frac{g_S^2}{2}X_a^2\big(K^2\pm m_f m_{f'}\big)\nonumber\\
&+ G_f^{-1}(K) G_{f'}^{-1}(K)\bigg]^2 G_{f''}^{-2}(K),
\end{align}
while for vector ($X=V,$ upper sign) and axial-vector ($X=A$, lower sign)
\bea
\label{Eq:simple_det_fact_VA}
&&D^{(X_a)}=\bigg[\frac{g_V^4}{16}\big(X_a^2\big)^2 + G_f^{-1}(K) G_{f'}^{-1}(K)\\\nonumber
  &&\qquad+\frac{g_V^2}{2} \Big( X_a^2\big(K^2\mp m_f m_{f'}\big) - 2 (X_a\cdot K)^2\Big) \bigg]^2 G_{f''}^{-2}(K),
\eea
where
$(f,f',f'')=(u,d,s)$ for $a=1,2,$ $(f,f',f'')=(u,s,d)$ for $a=4,5$, and $(f,f',f'')=(d,s,u)$ for $a=6,7.$

Now, one just has to use the expressions \eqref{Eq:simple_det_mix_SP}-\eqref{Eq:simple_det_fact_VA} of the determinant in the effective potential \eqref{Eq:U_f} to obtain the fermionic contribution to the curvature mass, according to its definition \eqref{Eq:dhatM2_SPVA}. Simple derivation with respect to the remaining fields leads to the $Q=0$ limit of the integral \eqref{Eq:I_X_def}, evaluated in Appendix~\eqref{app:PiCalcDetails}. For example, the determinant $D^{(X_a)}$ given in Eq.~\eqref{Eq:simple_det_fact_VA} leads through Eq.~\eqref{Eq:diff_log_D} to $\Delta m^{2,(X)\mu\nu}_{ab}=2N_c g_V^2 I^{X,\mu\nu}(m_f,m_{f'}),$ in accordance with the formula given in Table~\ref{Tab:Pi_all} for the first three flavor indices. We multiplied by the number of colors, $N_c$, as the determinant was calculated only in Dirac and flavor spaces.

\section{Decomposition of the self-energy tensor \label{app:decomp}}

In this appendix we consider at zero and finite temperature the decomposition into physical modes of the one-loop fermionic contribution to the momentum-dependent self-energy tensor of massive vector and axial-vector bosons, generically denoted by $\Pi_{\mu\nu}(Q)$. Special interest is devoted to the curvature mass of the modes, obtained from the self-energy in the limit $Q\to0$, which at $T\ne0$ represents the limit $q_0\to 0$, followed by $\q\to 0.$

\paragraph{$T=0$ case.} The \emph{vacuum} self-energy $\Pi^{\mu\nu}_{\rm vac}(Q)$ can be decomposed as
\be
\Pi^{\mu\nu}_{\rm vac}(Q) = \Pi_{\rm vac, L}(Q){\rm P}_{\rm L}^{\mu\nu}(Q) + \Pi_{\rm vac, T}(Q){\rm P}_{\rm T}^{\mu\nu}(Q),
\ee
with the 4-longitudinal and 4-transverse projectors
\be
   {\rm P}_{\rm L}^{\mu\nu}(Q)=\frac{Q^\mu Q^\nu}{Q^2}\quad \textrm{and}\quad {\rm P}_{\rm T}^{\mu\nu}(Q)=g^{\mu\nu}-{\rm P}_{\rm L}^{\mu\nu}(Q),
\label{Eq:PLT_zeroT}   
\ee
satisfying
\be
   {\rm P}_{\rm L/T}\cdot {\rm P}_{\rm L/T}={\rm P}_{\rm L/T},\ \ 
   {\rm P}_{\rm T/L}\cdot {\rm P}_{\rm L/T}=0, \ \ 
   {\rm P}_{\rm T\mu}^\mu=3{\rm P}_{\rm L\mu}^\mu=3.
\label{Eq:zT_prop_rel}
\ee
Writing the tree-level inverse propagator as $i{\cal D}^{-1}_{\mu\nu}(Q)=\hat m^2 {\rm P}^{\rm L}_{\mu\nu} + (\hat m^2 - Q^2){\rm P}^{\rm T}_{\mu\nu}$, where $\hat m^2$ is the tree-level curvature mass, one obtains from the Dyson equation $i{\cal G}^{-1}_{\mu\nu}(Q) = i{\cal D}^{-1}_{\mu\nu}(Q) + \Pi^{\rm vac}_{\mu\nu}(Q)$ the propagator
\be
   {\cal G}^{\mu\nu}(Q)=\frac{i {\rm P}_{\rm T}^{\mu\nu} }{-Q^2+\hat m^2 + \Pi_{\rm vac, T}(Q)} + \frac{i {\rm P}_{\rm L}^{\mu\nu}}{\hat m^2 + \Pi_{\rm vac, L}(Q)}.
\ee
It is evident that the curvature masses of the propagating (T) and nonpropagating (L) modes are:
\be
\hat M^2_{\rm vac, L/T}=\hat m^2 + \Pi_{\rm vac, L/T}(0).
\ee

In the $N_f=1$ case, due to the fermion number (current) conservation, the vector boson self-energy not only is transverse, that is, $Q_\mu \Pi^{\mu\nu}(Q)=0,$ but also satisfies $\Pi^{\mu\nu}(Q=0)\equiv 0,$ and therefore $\Pi_{\rm vac, L/T}(0)\equiv 0$, just like in the case of the photon polarization tensor in the QED. In the $N_f=2+1$ case the above relations hold due to Eq.~\eqref{Eq:Eq:CC_1-loop_final} for the vector boson self-energy with flavor indices listed in Eq.~\eqref{Eq:zero_div}. These indices correspond to the last three entries of Table~\ref{Tab:flavor_tr} (also for the first entry in the $\phi_3=0$ case), when the integrals involve fermion propagators with identical masses. For the first three entries of the table (except for the first one in the $\phi_3=0$ case) the vector polarization tensor is alike the axial-vector one, that is, $Q_\mu \Pi^{\mu\nu}(Q)\ne 0$ and $\Pi^{\mu\nu}(Q=0)\ne 0$, so that, using $\Pi_{\mu,{\rm vac}}^\mu(Q)=3 \Pi_{\rm vac, T}(Q) + \Pi_{\rm vac, L}(Q)$ and $\Pi^{\mu\nu}_{\rm vac}(Q=0)\propto g^{\mu\nu}$, one obtains 
\be
\Pi_{\rm vac, T}(0)=\Pi_{\rm vac, L}(0)=\Pi^{00}_{\rm vac}(0)=-\Pi^{11}_{\rm vac}(0).
\label{Eq:vac_curvature_mass_VA_comp}
\ee
Therefore, one can write unambiguously
\be
\hat M^2_\textrm{vac}=\hat m^2 + \Pi_\textrm{vac}(0), \quad \textrm{with}\quad
\Pi_\textrm{vac}(0) = \frac{1}{4}\Pi^\mu_{\mu,{\rm vac}}(0).
\label{Eq:vac_curvature_mass_VA}
\ee

\paragraph{$T\ne0$ case.} In a Lorentz-covariant formulation one has to take into account besides $Q^2$ a second Lorentz-invariant quantity, $\omega:=Q\cdot u$, where $u_\mu$ is the 4-velocity of the thermal bath, which satisfies $u^2=1.$ The self-energy depends on two scalars, $\omega$ and $q:=((Q\cdot u)^2-Q^2)^{1/2},$ which are interpreted as the Lorentz-invariant energy and modulus of the 3-momentum \cite{Weldon:1982aq}. In the rest frame of the thermal bath, also used here in what follows, one has $u^\mu=(1,{\bf 0})$, and therefore $\omega=q_0$ and $q=|\q|.$

A basis for the decomposition of self-energy has to be constructed from the four rank-2 tensors $g_{\mu\nu},$ $Q_\mu Q_\nu,$ $u_\mu u_\nu,$ and $Q_\mu u_\nu + Q_\nu u_\mu.$ The physically motivated basis \cite{Buchmuller:1992rs} consists of ${\rm P}^{\mu\nu}_{\rm L}$, given in Eq.~\eqref{Eq:PLT_zeroT}, and the three tensors
\begin{equation}
  \begin{aligned}
    {\rm P}_{\rm l}^{\mu\nu}&=\frac{u_{\rm T}^\mu u_{\rm T}^\nu}{u_{\rm T}^2} = -\frac{Q^2}{\q^2} u_{\rm T}^\mu u_{\rm T}^\nu,\\
    {\rm P}_{\rm t}^{\mu\nu}&=g^{\mu\nu} - {\rm P}_{\rm L}^{\mu\nu} - {\rm P}_{\rm l}^{\mu\nu}=-g^{\mu i}\left(\delta_{ij}-\frac{q_i q_j}{\q^2}\right)g^{j\nu},\\
    {\rm C}^{\mu\nu}&=\frac{Q^\mu u_{\rm T}^\nu + Q^\nu u_{\rm T}^\mu}{\sqrt{(Q\cdot u)^2 - Q^2}},
  \end{aligned}
\end{equation}
where $u_\mu^{\rm T}=u_\mu - (Q\cdot u) Q_\mu/Q^2$. ${\rm P}_{\rm t/l}$ are three-dimensional transverse/longitudinal projectors (they are both four-dimensional transverse, ${\rm P}_{\rm l}+{\rm P}_{\rm t}={\rm P}_{\rm T}$), while ${\rm C}_{\mu\nu}$ is not a projector. Further relations of interest, in addition to those in Eq.~\eqref{Eq:zT_prop_rel}, are
\begin{equation}
  \begin{aligned}
     & {\rm P}_{\rm l/t}\cdot {\rm P}_{\rm l/t} ={\rm P}_{\rm l/t}, \ \qquad \qquad \quad \ {\rm P}_{\rm l/t}\cdot {\rm P}_{\rm t/l}=0, \\
     &{\rm P}_{\rm t/L}\cdot {\rm P}_{\rm L/t}={\rm P}_{\rm l/L}\cdot {\rm P}_{\rm L/l}={\rm P}_{\rm t}\cdot {\rm C}={\rm C}\cdot {\rm P}_{\rm t} = 0,\\
     &{\rm P}_{\rm l}\cdot {\rm C}\cdot {\rm P}_{\rm l} = {\rm P}_{\rm L}\cdot {\rm C}\cdot {\rm P}_{\rm L} = 0,\ \ \ {\rm C}^2=-{\rm P}_{\rm L}-{\rm P}_{\rm l}, \ \ \\
    &{\rm C}\cdot{\rm P}_{\rm L} = {\rm P}_{\rm l}\cdot {\rm C}, \quad  \qquad\quad \quad \quad \ {\rm P}_{\rm L}\cdot {\rm C} = {\rm C} \cdot {\rm P}_{\rm l},\\
     &{\rm C}\cdot{\rm P}_{\rm L} + {\rm P}_{\rm L}\cdot {\rm C} = {\rm P}_{\rm l}\cdot {\rm C} + {\rm C} \cdot {\rm P}_{\rm l} = {\rm C},\\
     &d_{\rm l}:={\rm P}_{\rm l\mu}^\mu=1,\ \ d_{\rm t}:={\rm P}_{\rm t\mu}^\mu=2, \ \ {\rm C}_\mu^\mu = 0.
  \end{aligned}
  \label{Eq:nzT_tensor_rel}
\end{equation}

In the above basis the general self-energy tensor reads
\be
\Pi^{\mu\nu}(Q) =\sum_\textrm{p=t,l,L}\Pi_{\rm p}(Q){\rm P}_{\rm p}^{\mu\nu}(Q) + \Pi_{\rm C}(Q){\rm C}^{\mu\nu}(Q),
\label{Eq:nzT_tensor_decomp}
\ee
with tensor components given by ($d_t=2d_{\rm l/L}=2$)
\be
\Pi_{\rm p}=\frac{1}{d_{\rm p}}\tr(\Pi\cdot {\rm P}_{\rm p}), \quad \Pi_{\rm C}=-\frac{1}{2}\tr(\Pi\cdot {\rm C}).
\label{Eq:PiVA_comp}
\ee

The expression of the dressed propagator ${\cal G}_{\mu\nu}$ can be obtained with the method described in Ch.~5.2.2 of Ref.~\cite{leBellac}. Using a similar decomposition for ${\cal G}_{\mu\nu}$ as in Eq.~\eqref{Eq:nzT_tensor_decomp} and the Dyson equation $i{\cal G}^{-1}_{\mu\nu}(Q) = i{\cal D}^{-1}_{\mu\nu}(Q) + \Pi_{\mu\nu}(Q)$ in the identity ${\cal G}^{-1}_{\mu\nu}{\cal G}^{\nu\alpha}=g_\mu^\alpha$, one obtains by exploiting Eqs.~ \eqref{Eq:zT_prop_rel} and \eqref{Eq:nzT_tensor_rel}
\bea
    && {\cal G}^{\mu\nu}(Q)=\frac{i {\rm P}_{\rm t}^{\mu\nu} }{-Q^2+\hat m^2 + \Pi_{\rm t}(Q)} + \frac{i {\rm P}_{\rm l}^{\mu\nu}\big(\hat m^2 + \Pi_{\rm L}(Q)\big)}{\delta(Q)} \nonumber  \\
    && +\frac{i\big(-Q^2+ \hat m^2 + \Pi_{\rm L}(Q)\big){\rm P}_{\rm L}^{\mu\nu}}{\delta(Q)} - \frac{i\Pi_{\rm C}(Q) {\rm C}^{\mu\nu}}{\delta(Q)},
\eea
with $\delta(Q)=\big(\hat m^2 + \Pi_{\rm L}(Q)\big) \big(-Q^2+ \hat m^2 + \Pi_{\rm l}(Q)\big)+\Pi^2_{\rm C}.$ We will see below that $\Pi_{\rm C}(0,{\bf 0})\equiv 0,$ and hence the squared curvature masses of the remaining modes simplify to
\bea
\hat M^2_{\rm l/t/L}=\hat m^2 + \Pi_{\rm l/t/L}(0,{\bf 0}),
\eea
where $\hat m^2$ is the classical curvature mass squared and
\be
\Pi_{\rm p}(0,{\bf 0})=\Pi_{\rm vac}(0) + \Pi^{\rm mat}_{\rm p}(0,{\bf 0}),\quad {\rm p=l,t,L},
\ee
with the {\it vacuum} part $\Pi_{\rm vac}(0)$ defined in Eq.~\eqref{Eq:vac_curvature_mass_VA} and $\Pi^{\rm mat}_{\rm p}$ being the {\it matter} part.

The tensor components $\Pi^{\rm mat}_{\textrm{l/t/L/C}}(Q)$ can be given in terms of the Lorentz components of the self-energy tensor. Specifically for $q_0=0$, using the explicit expression of the projectors, one obtains from Eq.~\eqref{Eq:nzT_tensor_decomp}
\bea
&&\Pi_{\rm l}^{\rm mat}(0,\q)=\Pi_{00}^{\rm mat}(0,\q),\quad  \Pi_{\rm L}^{\rm mat}(0,\q)=-\frac{q_i q_j}{\q^2}\Pi_{ij}^{\rm mat}(0,\q), \nonumber\\ 
&&\Pi_{\rm C}^{\rm mat }(0,\q)=\frac{q_i}{|\q|}\Pi^{\rm mat}_{0i}(0,\q),
\label{Eq:comp_genSE}
\eea
while taking the trace in Eq.~\eqref{Eq:nzT_tensor_decomp} gives
\be
\Pi_{\rm t}^{\rm mat}(Q)=\frac{1}{2}\big[\Pi^\mu_{\mu, \rm mat}(Q)-\Pi_{\rm l}^{\rm mat}(Q)-\Pi_{\rm L}^{\rm mat}(Q)\big].
\label{Eq:trace_genSE}
\ee

The vector boson self-energy is 4-transverse ($Q_\mu \Pi^{\mu\nu}(Q)=0$) in the $N_f=1$ case and also in the $N_f=2+1$ case, for those flavor indices for which the bubble integral involves propagators of equal masses (for $\phi_3\ne0$, the last three lines of Table~\ref{Tab:flavor_tr}). In these cases $\Pi_{\rm L}(Q)=\Pi_{\rm C}(Q)\equiv 0$ and the 00 component of Eq.~\eqref{Eq:nzT_tensor_decomp} gives $\Pi_{00}^{\rm mat}(Q)=-\frac{\q^2}{Q^2}\Pi_{\rm l}^{\rm mat}(Q)$. From this relation or the first entry of Eq.~\eqref{Eq:comp_genSE} and from Eq.~\eqref{Eq:trace_genSE} one has
\be
\Pi_{\rm l}^{\rm mat}(0,{\bf 0})=\Pi_{00}^{\rm mat}(0,{\bf 0}), \quad
\Pi_{\rm t}^{\rm mat}(0,{\bf 0})=-\frac{3}{2}\Pi_{11}^{\rm mat}(0,{\bf 0}).
\label{Eq:nzT_PiV_comp_Lor-ind_rel_eq_mass}
\ee

For the axial-vector bosons and, in the $N_f=2+1$ case, for the vector boson self-energy involving bubble integrals with different fermion masses (for $\phi_3\ne0$, the first three lines of Table~\ref{Tab:flavor_tr}), the direct calculation presented in Appendix~\ref{app:prop_self_energy} shows that
\begin{subequations}
\bea
\Pi_{0i}^{\rm mat}(0,\q)&\equiv&0, \label{Eq:PiVA_rel1}\\
\underset{\q\to0}{\lim}\frac{q_i q_j}{q^2}\Pi_{ij}^{\rm mat}(0,\q)&=&\Pi_{11}^{\rm mat}(0,{\bf 0}). \label{Eq:PiVA_rel2}
\eea
\label{Eq:PiVA_rels}
\end{subequations}
As a result $\Pi_{\rm C}^{\rm mat}(0,\q)=0$ and, therefore, 
\be
\Pi_{\rm t/L}^{\rm mat}(0,{\bf 0})=-\Pi_{11}^{\rm mat}(0,{\bf 0}),\quad \Pi_{\rm l}^{\rm mat}(0,{\bf 0})=\Pi_{00}^{\rm mat}(0,{\bf 0}).
\label{Eq:nzT_PiA_comp_Lor-ind_rel}
\ee

\section{Proof of some properties of $\Pi_{\mu\nu}^{\rm mat}(Q)$\label{app:prop_self_energy}}

The one-loop \hbox{(axial-)}vector boson self-energy \eqref{Eq:SE_Nf3} can be given (see Table~\ref{Tab:flavor_tr}) in terms of the integral defined in Eq.~\eqref{Eq:I_X_def}, containing different fermion masses $m_f$ and $m_{f'}$, or (a linear combination of) its equal mass limit \eqref{Eq:I_X_one_mass}. In order to prove Eq.~\eqref{Eq:PiVA_rels}, it is enough to consider the integral in Eq.~\eqref{Eq:V_A_2masses}, obtained from Eq.~\eqref{Eq:I_X_def} by doing the Dirac trace. The following calculation refers to the \emph{matter} part, as indicated by the separation of the $k_0$ integral, which becomes a Matsubara sum in the imaginary time formalism.

\paragraph{Proof of \eqref{Eq:PiVA_rel1}}

Setting $q_0=0$ in \eqref{Eq:V_A_2masses}, one has
\be
  I^{V/A,0i}(0,\q; m_f, m_{f'})= i\int_{-\infty}^{\infty}dk_0\int_{\k}\frac{k_0(2 k_i - q_i) G_f(K)}{k_0^2-E_{\k-\q,f'}^2},
\ee
with $E_{\k,f}=(\k^2 + m_f^2)^{1/2}$ and $G_f^{-1}=K^2-m_f^2$. The integral is obviously zero, as the integrand is odd in $k_0$.

\paragraph{Proof of \eqref{Eq:PiVA_rel2}}
On the one hand, it follows again from Eq.~\eqref{Eq:V_A_2masses} by $q_0=0$ substitution that
\bea
&&\frac{q_i q_j}{\q^2}I^{V/A,ij}(0,\q;m_f,m_{f'})
\nonumber \\
&&=-i\int_{-\infty}^{\infty}dk_0 \int_{\k}\frac{\pm m_f m_{f'} - K^2 +\k\cdot\q-2(\k\cdot\q)^2/\q^2}{(k_0^2-E_{\k,f}^2)(k_0^2-E_{\k-\q,f'}^2)}.\nonumber \\
\eea
Changing to spherical coordinates in the $\k$ integral, and using $\k\cdot\q = k q \cos\vartheta\equiv k q x$ the $\q \to 0$ limit gives $-2k^2x^2$ in the numerator. Doing the angular integral leads to
\bea
 \lim_{\q\to 0} &&\frac{q_i q_j}{\q^2}I^{V/A,ij}(0,\q;m_f,m_{f'})=\nonumber\\
&& 4i\pi \int_{-\infty}^{\infty}dk_0\int_{0}^{\infty}dk\frac{\mp m_f m_{f'} + K^2 +k^2/3}{(k_0^2-E_{\k,f}^2)(k_0^2-E_{\k,f'}^2)}.\ \ 
\label{Eq:lim_pi_ij}
\eea

On the other hand, using the identity \eqref{Eq:rot_inv_3d} in Eq.~\eqref{Eq:V_A_2masses} gives
\be
  I^{V/A,11}(0,{\bf 0};m_f,m_{f'}) = i\int_K\frac{\mp m_f m_{f'} + K^2 +\k^2/3}{(k_0^2-E_{\k,f}^2)(k_0^2-E_{\k,f'}^2)}.
\ee
Changing to spherical coordinates in the $\k$ integral, the angular integral readily gives $4\pi$, as the integrand is independent of the angles. The resulting formula agrees with the rhs of Eq.~\eqref{Eq:lim_pi_ij}, from which Eq.~\eqref{Eq:PiVA_rel2} follows.

\end{document}